\documentclass{aa}

\usepackage{graphicx}
\usepackage{txfonts}
\usepackage{hyperref}
\usepackage{upgreek}
\let\originaleqref\eqref
\renewcommand{\eqref}{Eq.~\originaleqref}

\newcommand{\sectionname}{Sect.}

\usepackage[usenames]{color}  

\begin{document}

   \title{Generation of internal gravity waves by penetrative convection}
 

   \author{C. Pin\c con, K. Belkacem \and M. J. Goupil
          }

   \institute{LESIA, Observatoire de Paris, PSL Research University, CNRS, Universit\'e Pierre et Marie Curie,
   	Universit\'e Paris Diderot,  92195 Meudon, France\\
          \email{charly.pincon@obspm.fr}
             }

   \offprints{C. Pin\c con}
   \mail{charly.pincon@obspm.fr}
   \date{\today}
      
   \authorrunning{C. Pin\c con}

 
  \abstract
   {The rich harvest of seismic observations over the past decade provides evidence of angular momentum redistribution in stellar interiors that is not reproduced by current evolution codes. In this context, transport by internal gravity waves can play a role and could explain discrepancies between theory and observations.}
   {The efficiency of the transport of angular momentum by waves depends on their driving mechanism. While excitation by turbulence throughout the convective zone has already been investigated, we know that penetrative convection into the stably stratified radiative zone can also generate internal gravity waves. Therefore, we aim at developing a semianalytical model to estimate the generation of IGW by penetrative plumes below an upper convective envelope. The formalism is developed with the purpose of being implemented in 1D stellar evolution codes.
}
   {We derive the wave amplitude considering the pressure exerted by an ensemble of plumes on the interface between the radiative and convective zones as source term in the equation of momentum. We consider the effect of a thermal transition from a convective gradient to a radiative one on the transmission of the wave into the radiative zone. The plume-induced wave energy flux at the top of the radiative zone is computed for a solar model and is compared to the turbulence-induced one.}
   {
We show that, for the solar case, penetrative convection generates waves more efficiently than turbulence and that plume-induced waves can modify the internal rotation rate on shorter time scales. The result is solid since it holds despite a wide range of values considered for the parameters of the model. We also show that a smooth thermal transition significatively enhances the wave transmission compared to the case of a steep transition.}
   {Driving by penetrative convection must be taken into account as much as turbulence-induced waves for the transport of internal angular momentum. We propose a simple prescription that has the advantage of being easily implementable into 1D stellar evolution codes. We expect this mechanism to work in evolved stars, which will be subject to future investigations.}


   \maketitle

%
\section{Introduction}
%

Rotation is a fundamental ingredient of stellar evolution. It induces transport of angular momentum and of chemical elements that modifies the internal structure of stars \citep[e.g.,][]{Maeder2009}. For instance, rotationally induced mixing is able to refresh nuclear burning cores with hydrogen and thus to increase the star lifetime on the main sequence. Among other consequences, stellar age-dating is significantly affected by rotation \citep[e.g.,][]{Lebreton2014} such that the rotational history of stars has to be fully understood to obtain a relevant and complete picture of their evolution. 

This requirement is supported by an increasing number of observational facts. First, indirect constraints, such as anomalies in surface chemical abundances observed in stellar clusters, highlighted the importance of rotationally induced mixing \citep[see][for a review]{Charbonnel2008}. Second, the development of asteroseismology made direct measurements of the internal rotation profile possible. In the Sun, seismic measurements showed that its internal radiative zone rotates almost as a solid body  \citep[e.g.,][]{Brown1989,Garcia2007}. More recently, seismic data provided by the space-borne missions CoRoT \citep{Baglin2006a,Baglin2006b} and {\it{Kepler}} \citep{Borucki2010} enabled us to extend the study from the main sequence up to the red giant evolutionary phase. Thanks to the detection of mixed-modes \citep[e.g.,][]{Bedding2010,Mosser2012a}, it has been possible to estimate the core rotation rate of several subgiant stars observed by {\it{Kepler}} \citep{Deheuvels2012,Deheuvels2014} as well as the spinning down of the red giants core during their ascent of the vertical branch \citep{Mosser2012}. 

All these observational constraints emphasized the need for including transport of angular momentum as well as rotational mixing in stellar models. However, stellar evolution codes that take meridional circulation and shear-induced turbulence into account are unable to reproduce the observations by several orders of magnitude \citep{Eggenberger2012,Marques2013,Ceillier2013}. This suggests that other efficient transport processes must be included. 
Several mechanisms have already been investigated
such as the effects of magnetic fields \citep{Spruit2002,Heger2005,Cantiello2014,Rudiger2015} or mixed modes \citep{Belkacem2015b,Belkacem2015a}.
Internal gravity waves (hereafter IGW) can also play a role in the radiative zone of stars. 
They have been shown to be able to explain the nearly flat internal rotation profile observed in the Sun \citep{Zahn1997,Talon2002} and to be responsible for the cold side of the lithium dip observed in low-mass stars \citep{Talon1998b,Talon1998c,Talon2003,Talon2004,Talon2005}. 

The efficiency of the transport
by IGW crucially depends on the excitation mechanism.
IGW are preferentially generated at the boundary between the convective and radiative regions by convective motions.
\cite{Press1981} considered turbulent pressure at the base of the convective zone as driving process. 
He found that the wave energy flux could be expressed as the product of the mechanical convective energy flux at the base of the convection region with the ratio of the wave impedances in both regions (i.e., in the convective and the radiative regions).
\cite{Garcia1991} followed a similar approach but considered a distribution of convective eddies.
They assumed the turbulence to follow a Kolmogorov spectrum and took into account the incoherent behavior of the eddies.
Later, \cite{Zahn1997} generalized this model to a continuous wave spectrum. Finally, \mbox{\cite{Kumar1999}} adapted the work of \cite{Goldreich1994}
to the case of the excitation of IGW. In this model, IGW are generated by Reynold's stress throughout the convective zone
and the approach has the advantage of clearly taking into account the spatial and temporal correlations between the turbulent stochastic source and the waves. 
Most of the estimates of the transport by IGW used the latter formulation to include the effects of IGW in stellar radiative zones \citep[e.g.,][]{Talon2005,Charbonnel2013,Mathis2013,Fuller2014}. 

However, penetration of convective plumes into stably stratified layers can also generate IGW. Indeed, in the penetration region, plumes are decelerated by buoyancy braking and can release a part of their kinetic energy into waves. In the geophysical context, this mechanism has already been observed in laboratory experiments and investigated theoretically for atmospheric flows \citep{Townsend1966,Stull1976}. In astrophysics, excitation of internal waves by penetrative convection has been investigated by means of  2D numerical simulations \citep{Hurlburt1986,Andersen1994,Dintrans2005,Rogers2006a,Rogers2013}
and more recently, in 3D, spherical numerical simulations for the Sun \citep{Brun2011,Alvan2014,Alvan2015}. Nevertheless, it is difficult today to extrapolate numerical results to stellar regimes. For instance, numerical constraints require higher thermal diffusivities than in stellar interiors, resulting in much more important convective energy flux \citep{Rempel2004,Dintrans2005}
and so an expected overestimated wave energy flux. 

In this work, we propose a complementary approach
and elaborate a semianalytical model of excitation of IGW by penetrative convection. We consider the impact of the plumes on the stably stratified layers as the driving mechanism. 
The effect of the thermal transition on the transmission of the wave into the radiative zone is taken into account at the interface between the radiative and convective regions (hereafter, radiative/convective interface).
Our goal is to estimate how efficient is this mechanism compared to the excitation by turbulence as proposed by \cite{Kumar1999}.
If it is shown to be as efficient or more than the latter, such a simplified description will enable us to account for the transport of angular momentum by plume-induced waves in 1D stellar evolution codes.

The article is organized as follows. In \sectionname{} \ref{theoretical formalism}, we introduce the theoretical formalism. \sectionname{} \ref{plume description} presents the characteristics of the plumes and describes buoyancy braking process in the penetration zone. In \sectionname{} \ref{generation of IGW}, we summarize the method used to derive the wave energy flux and the main results are given. The wave energy flux is computed for a solar model in \sectionname{} \ref{application on a solar model}. The results are discussed and compared to those obtained with the formalism of \cite{Kumar1999} in \sectionname{} \ref{discussion}. Conclusions are formulated in \sectionname{} \ref{conclusion}.

%
\section{Theoretical formalism}
%
\label{theoretical formalism}

Our objective is to estimate the amount of plume kinetic energy 
transferred into wave energy in the penetration region.
For this purpose, we introduce the theoretical formalism and general concepts regarding IGW and spectral analysis. 
In the following, we will deal with a stationary random process of excitation by an ensemble of plumes and will use a statistical approach.

\subsection{Wave equation and source term}
\label{wave equation and source term}

The determination of the wave energy flux requires the calculation of the wave amplitude, which depends 
on the excitation mechanism. As convective plumes penetrate into the stably stratified medium, they perturbate the equilibrium state and generate waves by exerting pressure on the radiative/convective interface.
We assume that the total velocity field can be split
into two components
: the wave velocity field $\vec{v}(\vec{r},t)$, 
and the plume velocity field $\vec{V}_p(\vec{r},t)$. This actually corresponds to neglecting the dynamical effect of the turbulence inside the plumes.
Moreover, we assume that the plumes evolve independently of the wave velocity
and that there is no feedback from the waves on the plumes. This also suggests that the plumes and the waves follow their own continuity equation independently of each other. 
This is actually a good approximation if the wave velocity field is much smaller than the plume velocity field $|\vec{v}| \ll |\vec{V}_p|$, which will be verified a posteriori.

Under all this set of approximations, we will then focus our attention on the generation of waves by the stress exerted by the convective plumes, 
represented by $\boldsymbol{\nabla} \cdot (\rho \vec{V}_p\otimes \vec{V}_p)$ as the source term in the momentum equation, with $\rho$ the density at the equilibrium, $\boldsymbol{\nabla}$ the nabla operator and ($\otimes$) the tensorial product. For sake's of simplicity, we neglect any effects of rotation and adopt the Cowling approximation. Therefore, the linearized equations of momentum and of continuity with respect to the equilibrium state for the wave perturbations read 
\begin{align}
&\frac{\partial \vec{v}}{\partial t} +\frac{\boldsymbol{\nabla} p^\prime}{\rho} - \frac{\rho^\prime}{\rho} \vec{g} = -\frac{1}{\rho} \boldsymbol{\nabla} \cdot (\rho \vec{V}_p\otimes \vec{V}_p)\label{momentum eq}\\
&\frac{\partial}{\partial t} \left( \frac{\delta \rho}{\rho}\right) +\boldsymbol{\nabla} \cdot \vec{v} =0 \mbox{  ,} \label{continuity eq}
\end{align}
where $p^\prime$ and $\rho^\prime$ denotes Eulerian perturbations of pressure and density, $\delta \rho$ is the Lagrangian perturbation of density and $\vec{g}$ the gravitational acceleration at equilibrium. 

\subsection{Spectral density of the wave specific energy}

In the following, we want to determine the spectral distribution of the wave energy flux at each level in the star, generated by a population of penetrative plumes.

\subsubsection{Stationarity and ergodicity}

\begin{figure}
\centering
\includegraphics[scale=0.35, trim= 0cm 0cm 0cm 0cm,clip]{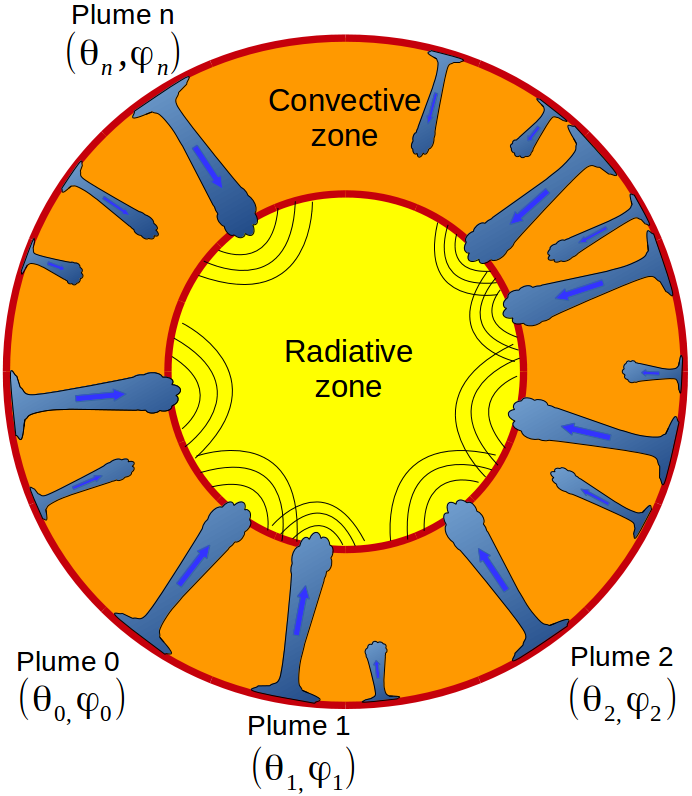} 
\caption{Schematic view of the star. The convective plumes occur in the upper layers of the star and go deeper into the convective region. They 
grow by turbulent entrainment of matter at their edges and reach the top of the radiative zone. There, each of them, characterized by their angular position $(\theta_i,\phi_i)$, releases a part of their kinetic energy and generate internal waves which can propagate towards the center. In our work, we suppose that there are no reflected waves propagating from the center to the radiative/convective interface.}
\label{schema star}
\end{figure}

We assume that at each time, an ensemble of several uniformly distributed incoherent plumes generate waves by penetrative convection (see \figurename{} \ref{schema star}, for a schematic representation). The process of excitation is then supposed to be random, stationary and ergodic,
and a statistical approach is valid since the number of plumes is high enough.
Hence, we can define the mean specific wave energy at a point $\vec{r}$ as
\begin{equation}
\left \langle E \right \rangle(\vec{r})
=\lim_{T\rightarrow +\infty}\frac{1}{T} \int_{-\infty}^{+\infty} \rho(r) |\vec{v}_T(\vec{r},t)|^2 \mathrm{d} t \mbox{  ,}
\label{mean wave energy}
\end{equation}
where $\vec{v}_T(\vec{r},t)$ is equal to the truncated part of the wave velocity field in the interval $[-T/2,T/2]$ and is null outside. The Parseval-Plancherel's theorem allows us to write
\begin{equation}
\int_{-\infty}^{\infty}\left \langle |\vec{v}_T(\vec{r},t)|^2 \right \rangle \mathrm{d} t =
\frac{1}{2 \pi} \int_{-\infty}^{\infty} \left \langle |\hat{\vec{v}}_T(\vec{r},\omega)|^2 \right \rangle \mathrm{d} \omega \mbox{  ,}
\label{plancherel}
\end{equation}
where $\omega$ is the temporal radian frequency and the symbol ( $\hat{ {} }$ ) denotes the time Fourier transform\footnote{The time Fourier transform of a function $X(\vec{r},t)$ is defined as

\centering $\mathcal{TF}[X]=\hat{X}(\vec{r},\omega)=\int_{-\infty}^{\infty} X(\vec{r},t) e^{-i\omega t} dt$  .}.
Then, using \eqref{plancherel} and since the integral and the limit commute, \eqref{mean wave energy} becomes
\begin{equation}
\langle E \rangle(\vec{r}) =\int_{-\infty}^{\infty} \epsilon (\vec{r},\omega) \mbox{ } \mathrm{d} \omega \mbox{  ,}
\end{equation}
where we have defined the spectral density of the wave specific energy
\begin{equation}
\epsilon (\vec{r},\omega)=\lim_{T\rightarrow +\infty}\frac{\rho(r)}{T} \frac{ \left\langle | \hat{\vec{v}}_T(\vec{r},\omega) |^2 \right\rangle}{2 \pi} \mbox{ .}
\label{spectral density}
\end{equation}
%

\subsubsection{Ensemble of $\mathcal{N}$ plumes}
\label{ensemble of plumes}
At each time, we assume that, on average, $\mathcal{N}$ identical plumes are penetrating into the stably stratified zone.
We note $(\theta_i,\phi_i)$ the angular position of the center of the plume $i$, that penetrates at the time $t_i$ with a velocity field $\vec{V}_{p,i}(\vec{r},t)$. Since the plumes are spatially and temporally well separated with each other, the source term in \eqref{momentum eq} can be rewritten as
\begin{equation}
\boldsymbol{\nabla} \cdot (\rho \vec{V}_p\otimes \vec{V}_p)=\sum_{i=1}^{+\infty} \boldsymbol{\nabla} \cdot (\rho \vec{V}_{p,i}\otimes \vec{V}_{p,i})\mbox{  .}
\end{equation}
Each plume $i$ generates a wave velocity field ${\vec{v}_i(\vec{r},t;\theta_i,\phi_i,\Delta t=t_i)}$ where $\Delta t$ denotes the time shift of the excitation event compared to the instant $t=0$. Given the linearity of the wave equation, the total wave velocity field is the superposition of all the contributions. In a similar way, at a fixed position $\vec{r}$, the truncated wave velocity field, $\vec{v}_T$, in the interval $[-T/2,T/2]$, must be the superposition of the wave velocity fields generated by a finite number $\mathcal{N}_{T}$ of plumes at different times\footnote{The choice of the plume ensemble depends on the position $\vec{r}$ since we have to consider the travel time of each plume-induced wave packet from its excitation site to this point $\vec{r}$.}
\begin{equation}
\vec{v}_T(\vec{r},t)=\sum_{i=1}^{\mathcal{N}_T} \vec{v}_i(\vec{r},t;\theta_i,\phi_i,\Delta t=t_i) \mbox{  .}\label{trunc velocity}
\end{equation}
When writing \eqref{trunc velocity}, we assume that the plume-induced wave packet $\vec{v}_i$ at any point $\vec{r}$ has a finite lifetime (otherwise, this would lead to an accumulation of wave energy in the radiative zone under the adiabatic hypothesis). Indeed, as suggested by the momentum equation \eqref{momentum eq}, it must be close to the characteristic plume lifetime; this is the consequence of the temporal correlation with the plume and the fact that no reflected wave is considered, but only propagative ones towards the center of the star. 

By the linearity and the time shift properties of Fourier transform,  we then obtain\footnote{We use the simplifying notation $\hat{\vec{v}}_i(\vec{r},\omega;\theta_i,\phi_i,\Delta t) =\hat{\vec{v}}_i(\Delta t)$. }
\begin{align}
 \left \langle | \hat{\vec{v}}_T (\vec{r},\omega) |^2 \right \rangle &=
\sum_{i=1}^{\mathcal{N}_{T}}\left\langle \left|\hat{\vec{v}}_i(\Delta t=0)\right|^2\right \rangle \nonumber \\
&+\sum_{i=1,j\ne i}^{\mathcal{N}_{T}}\left\langle  e^{-i\omega (t_i-t_j)}\hat{\vec{v}}_i(\Delta t=0)\overline{\hat{\vec{v}}}_j(\Delta t=0)\right \rangle \mbox{  ,}
\label{v_T ensemble}
\end{align}
where the bar denotes the complex conjugate. The statistical average covers the spatial and temporal distribution of the plumes. The plumes are incoherent with each other and so the phase lag between each component is randomly distributed. Note that the same assumptions were first used by \cite{Garcia1991}. As a consequence, the second term in the right-hand side of \eqref{v_T ensemble} vanishes.
Moreover, the plumes being uniformly distributed and independent with each other, the probability for the plumes $1,2,...,\mathcal{N}_T$ to be located in the solid angles $\mathrm{d}\Omega_1,\mathrm{d}\Omega_2,...,\mathrm{d}\Omega_{\mathcal{N}_T}$, respectively, is equal to $\prod_1^{\mathcal{N}_T} \mathrm{d}\Omega_i/ \left(4 \pi\right)^{\mathcal{N}_T}$. 
Thus, using \eqref{v_T ensemble}, \eqref{spectral density} becomes
\begin{equation}
\epsilon (\vec{r},\omega)=\lim_{T\rightarrow +\infty}\frac{1}{T} \frac{\rho(r)}{2 \pi}
\sum_{i=1}^{\mathcal{N}_{T}} \left\{\frac{1}{\left(4 \pi \right)^{\mathcal{N}_{T}}} \int_{\Omega_j} \left|\hat{\vec{v}}_i(\Delta t=0)\right|^2\prod_{j=1}^{\mathcal{N}_{T}}\mathrm{d}\Omega_j\right\} \label{epsilon average} \mbox{  .}
\end{equation}
Since we suppose that the plumes are identical and differ from each other only by their angular position, \eqref{epsilon average} becomes
\begin{equation}
\epsilon (\vec{r},\omega)=\lim_{T\rightarrow +\infty}\frac{\mathcal{N}_{T}}{T} \frac{\rho(r)}{8 \pi^2} \int_{\Omega_0} 
\left| \hat{\vec{v}}_0(\vec{r},\omega;\theta_0,\phi_0,\Delta t=0)\right|^2 \mathrm{d}\Omega_0 \mbox{  ,}
\label{epsilon one plume}
\end{equation}
where $\hat{\vec{v}}_0$ is the Fourier transform of the wave velocity field generated by one single plume with the angular position $(\theta_0,\phi_0)$ at $t_0=0$.

To go further, the plume destruction rate must equal the plume emerging rate in order to ensure a constant number of penetrating plumes over time.
If $\tau_p$ denotes the characteristic plume lifetime and if it is supposed to be the same for all of them, this rate is equal to $\mathcal{N}/ \tau_p$. In other words, over a time $T \gg \tau_p$, a total number of $\mathcal{N}_T\sim\mathcal{N} T/ \tau_p$ plumes occur and contribute to the wave energy in \eqref{epsilon one plume}.
Finally, the spectral density of specific wave energy converges on average to 
\begin{equation}
\epsilon (\vec{r},\omega)= \mathcal{N}\frac{\nu_p}{ 2 \pi }\rho(r)
\int_{\Omega_0} 
\left| \hat{\vec{v}}_0(\vec{r},\omega;\theta_0,\phi_0,t_0=0)\right|^2 \frac{\mathrm{d}\Omega_0}{4 \pi} \mbox{  ,}
\label{spectral density ensemble}
\end{equation}
where $\nu_p=1/\tau_p$. 

We first notice that the spectral density of wave specific energy is proportional to the number of plumes, which is the consequence of their incoherent behavior
(in the case of coherent plumes, the wave specific energy would be proportional to $\mathcal{N}^2$). 
Second, we see that the use of a statistical approach simplifies the calculation since the case of an ensemble of several plumes is reduced to the study of the excitation by one single plume.

\subsection{Mean radial wave energy flux per unit of frequency}

For convenience, the Eulerian wave velocity field $\vec{v}_0$ is expanded onto the spherical harmonics
\begin{align}
\vec{v}_0(\vec{r},t)=\sum_{l,m} &\mathrm{v}_{r,l,m}(r,t) Y_l^m(\theta,\phi ) \mbox{ }\vec{e}_r+\mathrm{v}_{P,l,m}(r,t)\boldsymbol{\nabla} Y_l^m(\theta,\phi ) \nonumber \\
&+\mathrm{v}_{T,l,m}(r,t)\boldsymbol{\nabla} \wedge \left(Y_l^m(\theta,\phi )\mbox{ }\vec{e}_r\right) \mbox{  ,}
\label{decomposition wave velocity} 
\end{align}
where $(r,\theta,\phi)$ are the spherical coordinates, $l$ is the angular degree, 
$m$ the azimuthal number, $\vec{e}_r$ the radial unit vector, and where $\mathrm{v}_{r,l,m}$, $\mathrm{v}_{P,l,m}$ and $\mathrm{v}_{T,l,m}$ are the radial, poloidal and toroidal components of the wave velocity field, respectively.
Given the decomposition in \eqref{decomposition wave velocity}, \eqref{spectral density ensemble} averaged over the solid angle represents the angular distribution of energy and reads
\begin{align}
\mathcal{E} (r,\omega) &= \frac{1}{4 \pi} \int_{\Omega} \epsilon(r,\theta,\phi,\omega) \mathrm{d} \Omega \nonumber \\
&=\mathcal {N} \sum_{l,m} \widetilde{\mathcal{E}}_{l,m}(r,\omega) \label{spectral density (l,m)} \mbox{  ,}
\end{align}
where we have defined
\begin{equation}
\widetilde{\mathcal{E}}_{l,m}(r,\omega)= \frac{\nu_p}{8 \pi^2} \rho(r)\int_{\Omega_0} \left\{ \left|\hat{\mathrm{v}}_{r,l,m}\right|^2+l(l+1) \left|\hat{\mathrm{v}}_{h,l,m}\right|^2 
\right\} \frac{\mathrm{d}\Omega_0}{4\pi} \mbox{  ,} \label{spectral density term}
\end{equation}
with $\left|\hat{\mathrm{v}}_{h,l,m} \right|^2=\left|\hat{\mathrm{v}}_{P,l,m}\right|^2+\left| \hat{\mathrm{v}}_{T,l,m}\right|^2$. The mean radial wave energy flux by unit of frequency for an angular degree $l$ and an azimuthal number $m$ is thus obtained by multiplying the $(l,m)$ contribution to the spectral density of the specific wave energy with the radial group velocity $V_{gr}$ in a similar way to \cite{Zahn1997}
\begin{equation}
\mathcal{F}_r(r,\omega,l,m)=\mathcal{N}\hspace{0.1cm} \widetilde{\mathcal{E}}_{l,m}(r,\omega) \hspace{0.1cm} V_{gr}(r,\omega,l) \mbox{  .}
\label{flux (l,m)}
\end{equation}
Indeed, IGW are dispersive waves and the energy carried in the radial direction by such a wave packet travels at first order with the radial group velocity \citep[see][]{Lighthill1978} given by
\begin{equation}
V_{gr}(r,\omega,l)=\frac{\partial \omega }{\partial k_r}= \frac{\omega^2}{N^2} \frac{(N^2-\omega^2)^{1/2}}{k_h} \mbox{  ,}
\label{group_velocity}
\end{equation}
where $k_r$ is the local radial component of the wave vector \citep{Press1981}
\begin{equation}
k_r(r,\omega)=\left(\frac{N^2}{\omega^2}-1\right)^{1/2}k_h \mbox{  ,}
\label{k_r}
\end{equation}
with $k_h=\sqrt{l(l+1)}/r$ the horizontal component of the wave vector and $N$ the Brunt-Väisälä frequency.
The total mean radial wave energy flux per unit of frequency due to the penetration of several plumes is then obtained by summing all the $(l,m)$ components given by equation \eqref{flux (l,m)}.

%
\section{A simple description of convective plumes}
%
\label{plume description}

The computation of the wave energy flux requires the knowledge of the wave velocity field whose the amplitude directly depends on the excitation mechanism. We then need to model the driving process expressed by the plume-related term in \eqref{momentum eq}. For this purpose, we present here a simple physical description of convective plumes and their velocity profile in the penetration zone.

\subsection{Plume velocity field}
\label{plume velocity field}
The disturbance of the radiative/convective interface due to a single plume is localized in space and in time. Therefore, a plume is described by a characteristic radius $b$ and a characteristic lifetime $\tau_p$ in the penetration region.
We assume that the plume velocity field follows the Gaussian form\footnote{Under the assumption of a bell-shaped temporal profile, $\tau_p$ is defined as the time during which the plume kinetic energy is higher than about $60 \%$ of its maximal value.} proposed by \cite{Townsend1966}
and we also assume that the horizontal profile is maintained in the penetration region
\begin{equation}
\vec{V}_p(\vec{r})=V_0(r)\hspace{0.1cm} e^{-S_h^2/2b^2}e^{- t^2/\tau_p^2 }\hspace{0.1cm} \vec{e}_r \mbox{  ,}
\label{profil_Vp}
\end{equation} 
with
\begin{equation}
S_h(\vec{r};\theta_0,\phi_0)=r\arccos\left[\sin \theta_0 \sin \theta \cos(\phi-\phi_0)+\cos \theta \cos \theta_0\right] \mbox{  ,}
\label{def S_h}
\end{equation}
where $(\theta_0,\phi_0)$ are the angular coordinates of the center of the plume and $S_h$ corresponds to the distance on the sphere from the center of the plume.
%

\subsection{Plume radius}
\label{plume radius}

The plume radius at the bottom of the convective zone is estimated from the model of turbulent plumes by \cite{Rieutord1995}
who derived the expression 
\begin{equation}
b=\frac{z_0}{\sqrt{2}}\frac{3 \alpha_E(\Gamma_1-1)}{2\Gamma_1-1} \mbox{  ,}
\label{largeur_panache}
\end{equation}
where $z_0$ is the thickness of the convective zone, $\Gamma_1$ the adiabatic coefficient and $\alpha_E$ the entrainment coefficient whose value is usually taken as $0.083$ \citep{Turner1986}. In the solar case and for a monoatomic gas ($\Gamma_1=5/3$), we find $b\approx 10^4$ km, which is about five times smaller than the size of the biggest turbulent eddies at this radius as given by the MLT.

\subsection{Vertical velocity profile in the penetration zone}
\label{vertical velocity profile}

Once the plume has penetrated into the stably stratified layers, it is less dense than the surrounding medium and is slowed down by buoyancy braking. We adopt the description of \cite{Zahn1991} who considers a stationary flow in balance with buoyancy. Such an approach is justified by the high Péclet number at the base of the convective zone. 
\cite{Zahn1991} could estimate the penetration length $L_p$
and showed that it depends on a filling factor, the radiative diffusivity scale height, the total energy flux and the plume kinetic energy at the base of the convective envelope. 
In the solar case, he found $L_p\sim0.5 H_p \sim10^4$ km.
Nevertheless, using helioseismology, \cite{Basu1997} set an upper limit for the overshoot at $0.05H_p$, with $H_p$ the pressure scale height. In the framework of the model proposed by \cite{Zahn1991}, \cite{Basu1997} used stellar models with a simple adiabatic extent of the convective zone above a discontinuous thermal transition. In contrast, \cite{Dalsgaard2011} demonstrated that smoother transitions were also possible to explain seismic observations of the Sun. Such a kind of transition is supported by more realistic models of penetrative convection taking into account interaction with the upflows and a distribution for the velocities of the plumes \citep[e.g.,][]{Rempel2004}. Therefore, given the lack of knowledge and constraints about the penetration process, we will consider the penetration length as a parameter of the model whose value will be chosen around $0.1 H_p$.

Following the work of \cite{Zahn1991}, the vertical plume velocity $V_0$ in the penetration region is given by\footnote{We conserve the same velocity profile than in \cite{Zahn1991}, but we  consider the penetration length $L_p$ as a free parameter of the model.}
\begin{equation}
V_0(z)=V_b\left[1-\left(\frac{z}{L_p}\right)^2\right]^{1/3} \mbox{  ,}
\label{vitesse panache penetration}
\end{equation} 
where $z=r_b-r$ with $r_b$ the radius of the base of the convective zone as given by the Schwarzschild's criterion and $V_b$ the initial vertical plume velocity in the penetration region, at the radius $r_b$. To go further, we use the model of \cite{Rieutord1995}, so that
\begin{equation}
V_b=\left( \frac{8 F_{1p}}{\pi \rho_b b^2}\right)^{1/3} \mbox{  ,}
\label{V_b rieutord}
\end{equation}
where $F_{1p}$ represents the total luminosity (kinetic + enthalpic) carried by one single plume and $\rho_b$ is the mean density at the base of the convection zone. To estimate $F_{1p}$, we assume that each plume carries an energy equals, on average, to the one carried by turbulent eddies in the interplume medium such that
\begin{equation}
\mathcal{N} F_{1p}\approx L_{*} \mathcal{A} \mbox{  ,}
\end{equation}
where $L_{*}$ is the star luminosity and $\mathcal{A}=\mathcal{N}b^2/4r_b^2$ is the filling factor related to the fraction of the area occupied by the downdrafts at the base of the convective zone. Then, \eqref{V_b rieutord} becomes
\begin{equation}
V_b=\left( \frac{2 L_{*}}{\pi \rho_b r_b^2}\right)^{1/3} \mbox{  ,}
\label{V_b}
\end{equation}
which leads to $V_b \approx 185$~m~s${}^{-1}$ in the case of the Sun, i.e about 7 times higher than the convective velocity $\mathrm{v}_c \sim 25$~m~s${}^{-1}$ as given by the MLT at the base of the convective zone.

\subsection{Plume lifetime}
\label{plume lifetime}

The plume lifetime is certainly the more difficult plume characteristic to estimate. After having reached the base of the convective zone, the plume penetrates into the stably stratified zone. There,  it is slowed down and destroyed after a characteristic time $\tau_p$. 
For sake's of simplicity, we prefer here to tackle the issue by using orders of magnitude. We identify three potential processes working on the destruction of the plume:
\begin{enumerate}
\item
{\it{Radiative thermalization:}} in the penetration zone, the plume loses its identity due to radiative thermal diffusion on a time scale, $t_{rad}$. It can be estimated by
\begin{equation}
t_{rad} \sim \frac{L_p^2}{K_b} \mbox{  ,}
\end{equation}
where $K_b$ is the radiative conductivity at the top of the stably stratified layer. In the Sun, we find $t_{rad,\odot}\sim10^{11}$~s, so it is much longer than the dynamical time scale $t_{d,\odot} \sim L_p/V_b \sim$~10${}^4$s.  
As already mentioned, advection of adiabatic convective matter is faster than thermal exchange because of a high Péclet number at the base of the convective zone. Plume thermalization by radiative diffusion is thus inefficient, except in the transition region where the plume is slowed down enough for the dynamical time scale to be large enough. \\
\item {\it{Turbulence inside the plume:}}
although the plume is described like a coherent flow with well-defined radius and vertical profile, it is in fact turbulent at small scales as explained in \cite{Montalban2000} who suppose that turbulence is generated by shear flow in the penetration zone. In this case, the plume lifetime should be equal to a few turnover time scales of the biggest turbulent eddies whose the velocity and the size are about the ones of the plume, $V_b$ and $b$, respectively. The turbulent time scale, $t_{turb}$, can be then approximated by
\begin{equation}
t_{turb}\sim \frac{b}{V_b} \mbox{  .}
\end{equation}
For the Sun, we find $t_{turb,\odot} \sim 10^{5}$ s, which is much lower than the radiative time scale, but still one order of magnitude higher than dynamical time scale. Note that this value is close to the convective turnover time scale as predicted by the MLT, i.e., $t_{c} \sim 10^6$ s at the base of the solar convective envelope.\\
\item {\it{Restratification by lateral baroclinic eddies:}}
finally, it is worth considering the restratification phenomenon observed in the terrestrial oceans \citep{Jones1997} as a potential process operating in the stellar penetration region.
In most locations in the oceans, the surface layers are stably stratified. However, convection  driven by cooling at the sea-surface can occur in some particular region. Then, convective plumes develop by turbulent entrainment of matter while going deep in the sea. When convection stops, the phase during which the plume is destroyed and loses its identity is called restratification. At this point, the density gradient between the plume and the surrounding stable medium generates a thermal wind. This latter is subject to a baroclinic instability creating lateral eddies able to transfer density between both mediums and thus able to homogenize the region. By adopting the model of \cite{Jones1997} to the solar case (see \appendixname{} \ref{restratification} for details), the restratification time scale is about 
$t_{res}\sim 10^6-10^7\mathrm{s}$, which is in the order of magnitude of the convective eddy turnover. 
\end{enumerate}
The plume lifetime is likely to be the consequence of all these potential processes. The simple abovementioned estimates give reasonable values in the range of $10^5-~10^7$~s for the Sun, i.e around the convective turnover time scale as given by the MLT. Thus, the plume lifetime will be considered as a parameter of the model whose the value will be chosen to be of the order of magnitude of the convective turnover time scale.

%
\section{Generation of IGW by penetrative convection}
%
\label{generation of IGW}

\subsection{Modeling the penetration zone}

\begin{figure}
\centering
\includegraphics[scale=0.2, trim= 0cm 0cm 0cm 0cm,clip]{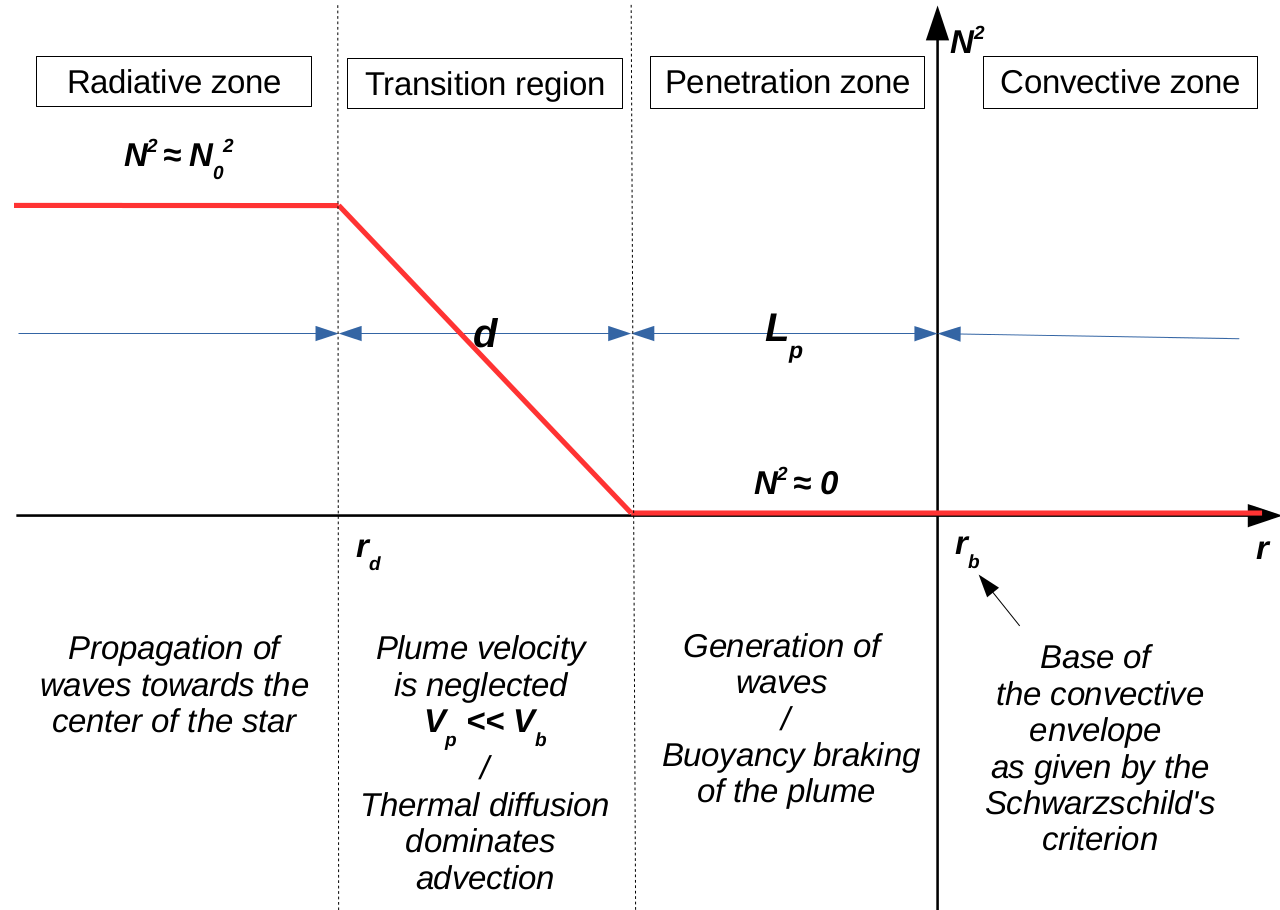} 
\caption{Modeling of the interface between the radiative zone an the convective zone.
$r_b$ corresponds to the radius at the base of the convective envelope as given by the Schwarzschild's criterion, $L_p$ is the penetration length over which the plume is decelerated and generates waves, $d$ denotes the extent of the transition region (where the plume velocity is negligible compared to $V_b$, its value at the entry of the penetration region) and $r_d$ is the radius at the top of the radiative zone.
}
\label{schema situation}
\end{figure}

Penetrative convection generates IGW at the radiative/convective interface and modifies the stellar equilibrium stratification. Therefore, we need to model this region to properly describe the excitation process. The scenario we consider in this work follows in part the description given by \cite{Zahn1991}. The situation is illustrated in \figurename{} \ref{schema situation}. The plume reaches the base of the convective zone located at the radius $r_b$ as prescribed by the Schwarzschild's criterion. By inertia, the column of fluid penetrates into the stably stratified zone over a penetration length $L_p$. The density contrast between the plume and the surrounding medium causes buoyancy braking and the plume slows down, releasing a part of its kinetic energy into waves. Indeed, as explained by \cite{Zahn1991}, since the Péclet number at the bottom of the convective zone is very high (${\rm P_e}\sim10^5 - 10^6$), advection of convective matter is more efficient than thermal exchanges so that the plume keeps its identity. Then, it imposes a quasi-adiabatic stratification in the penetration zone, resulting in a nearby vanishing Brunt-Väisälä frequency $N^2 \approx 0$.
Once its velocity is small enough (i.e., much smaller than its value at the entry of the penetration zone, $V_b$), the Péclet number decreases and the thermalization of the plume's material occurs in a thermal transition region; there, the temperature gradient switches from quasi-adiabatic one to radiative one over a distance $d$. In a first approach, we merely suppose that the Brunt-Väisälä frequency $N^2$ increases linearly to $N_0^2$, the value at the top of the radiative zone, as in the work of \cite{Lecoanet2013}. By this way, we aim at investigating how the plume-induced wave energy depends on the steepness of the transition. 

%

\subsection{Methodological approach}

We now have to derive the time Fourier transform of the wave velocity field which leads to the spectral description of the wave energy flux through \eqref{spectral density}. For this purpose, we take the time Fourier transform of \eqref{momentum eq} and \eqref{continuity eq}. We also assume that the oscillations are adiabatic so as to close the system. The first-order system of two differential equations obtained can be rewritten in the following form (see \appendixname{} \ref{adiabatic wave equation} and \ref{modelling stress} for a detailed derivation)
\begin{equation}
\frac{d\vec{z}}{d r}(r,\omega)=\vec{A}(r,\omega)\vec{z}(r,\omega)+\vec{b}(r,\omega)
\label{differential system z}
\end{equation}
where we have defined
\begin{equation}
\vec{z}=\left( \begin{array}{c}
\hat{\mathrm{v}}_{r,l,m} \\
\hat{\mathrm{v}}_{h,l,m} 
\end{array} \right)
\mbox{  ,}
\hspace{0.6cm}
\vec{b}=\left( \begin{array}{c}
0 \\
i F_2/r\omega
\end{array} \right)\mbox{  ,} \label{z and b}
\end{equation} 
\begin{equation}
\vec{A}=\left( \begin{array}{cc}
2/r-g/c^2 & \left( S_l^2 - \omega^2\right)r/c^2 \\
\left(\omega^2-N^2\right)/ r\omega^2  & -d_r\ln\left(\rho r\right)-g/c^2
\end{array} \right)
\label{matrix A}\mbox{  ,}
\end{equation}
where $g$ is the gravitational acceleration, $c$ is the sound speed, $S_l$ is the Lamb frequency and $d_r \equiv d/dr$ is the derivative with respect to $r$. The term $F_2$
is defined by \eqref{F2}; it contains information on the spatial and temporal correlations between the wave and the plume, and we assume that it only exists in the penetration region where the wave driving is stronger. In this region, the inhomogeneous equations are solved using the method of variation of parameters and using the plume description presented in \sectionname{} \ref{plume description}.
In the convective and radiative zones, we use the WKB asymptotic solution for the homogeneous differential system, which is valid in these regions contrary to the transition region where Airy functions are considered.

To go further, we impose the continuity of the radial and horizontal displacements at the top of the radiative zone in $r_d$ and at the beginning of the transition region in $r_d + d$. For boundary conditions, we first consider a pure regressive wave propagating towards the center in the radiative zone. That means we do not consider the possible reflection of the wave in the center of the star and so do not allow for standing waves to establish. Finally, in the convective zone, we consider a pure evanescent wave which is generated in the penetration zone and is damped towards the surface.

\subsection{Wave energy flux}
\label{wave energy flux main}
In the following, we summarize the main results of the detailed calculation given in \appendixname{} \ref{derivation of the wave energy flux}. The mean radial wave energy flux per unit of frequency generated by $\mathcal{N}$ plumes for an angular degree $l$ and an azimuthal number $m$ reads
\begin{equation}
\mathcal{F}_r(r,\omega,l,m)=f(\gamma_d) \frac{\mathcal{N}}{4} \frac{\sqrt{l(l+1}}{4 \pi r^2}\frac{\left(N_0^2-\omega^2\right)^{1/2}}{N_0^2} \frac{e^{-\omega^2/4 \nu_p^2}}{\nu_p} \mathcal{B}_l \mathcal{H}_l^2 \mbox{  ,} 
\label{spectral wave energy flux}
\end{equation}
with
\begin{align}
\mathcal{H}_l&= \frac{1}{4\pi}\int_{r_b-L_p}^{r_b} \frac{1}{\rho}\frac{d}{dr}\left( \rho V_0^2\right)
\rho^{-1/2} r^{-3/2} \left( \frac{r_b-L_p}{r}\right)^{\Lambda} \mathrm{d}m \label{H_l radial}\\
\mathcal{B}_l&=\frac{1}{4 \pi} \int_{\Omega_0} \left| \beta_l^m\right|^2 \mathrm{d} \Omega_0
\label{horizontal average}\\
\beta_l^m&(\theta_0,\phi_0)=\int_{\Omega} e^{-s_b^2/b^2} \overline{Y_l^m}\mathrm{d} \Omega \label{beta l m}\\
\gamma_d&=\left(\frac{k_h d N_0}{\omega}\right)^{2/3} \frac{N_0^2-\omega^2}{N_0^2} \sim \left( k_r d\right)^{2/3} \label{gamma_d}\\
f(x)&=\left \lbrace
\begin{array}{cl}
1  &  \mbox{if $x < 1$} \\
D \sqrt{x} \left(1-\omega^2/N_0^2\right)^{-1}&  \mbox{if $x > 1$}
\end{array}
\right. \label{transition function}
\end{align}
where $\Lambda=\sqrt{l(l+1)}$, $\mathrm{d} m=4\pi \rho r^2 \mathrm{d} r$, $D \approx 3.7$ a numerical factor, $s_b=S_h(r_b,\theta,\phi;\theta_0,\phi_0)$ following \eqref{def S_h} and $d$ is the length of the transition region (see \figurename{} \ref{schema situation}) .
We can first notice that the spectral density of wave luminosity, $\mathcal{L}=4 \pi r^2 \mathcal{F}_r$, is conserved at each level in the star, which is the consequence of the adiabatic hypothesis for the waves in a non-rotating star. Now, let us discuss the different terms:
\begin{enumerate}
\item $\mathcal{H}_l$ corresponds to the wave driving term and it is representative of the instantaneous power injected into the wave in the whole penetration zone\footnote{Indeed, the integrand in \eqref{H_l radial} is the product of a first term related to the plume-induced vertical force per unit of mass with a second term proportional to the radial wave velocity field, solution of the homogeneous wave equation in the WKB approximation.}. It decreases with the angular degree $l$ since the decay length of an evanescent wave in the penetration zone scales as $\Lambda^{-1}$.  In our model, it is also the only term which depends on the penetration length through the domain of integration and the plume velocity. If $L_p$ decreases, the buoyancy braking is stronger and the excitation is more efficient since the plume deposits its energy where the evanescent wave has a higher amplitude. Inversely, if $L_p$ increases, the plume energy is transferred into wave on a longer distance where the wave amplitude decreases rapidly.

\item $e^{-\omega^2/4 \nu_p^2}/\nu_p$ represents the temporal correlation term. The width of the spectral envelope is around $\nu_p$, which means that the transit time at a point $\vec{r}$ of a wave packet generated by one single plume is around $1/\nu_p=\tau_p $, so that the assumptions made in \sectionname{} \ref{ensemble of plumes} are verified. An increase of the characteristic plume lifetime causes an increase of the spectrum amplitude but a decrease of the spectrum width. In other words, the plume energy is transferred into higher frequencies at the expense of the lower ones, but the total wave energy is conserved since the integration of the flux over $\omega$ does not depend on $\nu_p$.
This is the consequence of the hypothesis of stationarity for the driving mechanism.

\item $\beta_l^m$ expresses the horizontal spatial correlation between the wave and the plume.

\item $\mathcal{B}_l$ corresponds to the horizontal correlation term averaged over the angular position of the plume $(\theta_0,\phi_0)$. Since the plumes are supposed to be horizontally, uniformly distributed at each time, the spherical symmetry is conserved and $\mathcal{B}_l$ does not depend on the azimuthal order $m$. This term decreases with $l$ since the spherical harmonics presents more and more zeros with higher degrees. Moreover, a change of behavior is expected when the horizontal wavelength, $\lambda_h = 2 \pi r_b/\Lambda$, becomes smaller than plume radius, i.e., for a degree $l_{crit} \sim 2 \pi r_b/b$. In the case of small degrees $l\ll l_{crit}$, an approximated analytical expression in agreement with the numerical integration is given by
\begin{eqnarray}
\mathcal{B}_l \sim \frac{\pi}{4} \left(\frac{b}{r_b}\right)^4 e^{-l(l+1) b^2/2 r_b^2}  & \mbox{for $l\ll l_{crit}$} \mbox{  .} \label{B_l theorique}
\end{eqnarray}
Since the wave energy flux per unit of frequency is proportional to $\Lambda \mathcal{B}_l$ (neglecting the dependence of $\mathcal{H}_l$ with $l$), \eqref{B_l theorique} shows that the energy flux should be maximal for a degree $l_{max} \sim r_b/b$.

\item $f(\gamma_d)$ is a transmission function. It discriminates between the limiting case of a very sharp, almost discontinuous transition for $\gamma_d \ll1$ and the one of a smoother transition for $\gamma_d \gg1$. It physically corresponds to the ratio of the transmission coefficients in both cases. A very sharp transition has no effect on the wave dynamics. In this case, the flux is inversely proportional to the value of the Brunt-Väisälä frequency at the top of the radiative zone, $N_0$. That means that the higher is the step, the higher is the wave impedance in the radiative zone \citep{Press1981} and the smaller is the wave energy transferred.
The smoother case occurs when the distance of the transition, $d$, is in the same order of magnitude or larger than the radial wavelength, so that the transition has an impact on the waves and improves its transmission into the propagative zone.
In the special case of a linear transition profile, the transmission of the wave energy flux is enhanced by a multiplying factor $(k_r d)^{1/3}$ compared to the discontinuous case, in agreement with \cite{Lecoanet2013}. We highlight here the effect of the steepness of the transition on the flux; this latter depends on the slope of the Brunt-Väisälä frequency in the transition zone which plays the role of a pseudo-impedance 
\begin{equation}
\mathcal{F}_r \propto \left( \frac{d}{N_0^2}\right)^{1/3} \approx \left| \frac{d N^2}{dr}\right|^{-1/3} \mbox{  ,}
\end{equation}
so that the smoother the transition, the higher the transmission of the wave energy.
\end{enumerate}

\subsection{Simplified expression for the wave energy flux}

Finally, it is possible to derive an approximate expression for \eqref{spectral wave energy flux} in the whole radiative zone. This gives the advantage of expressing the result in terms of the characteristic physical quantities of the problem and it permits to avoid the numerical calculations of integrals. By using \eqref{B_l theorique} for $l\ll l_{crit}$ and the assumption  $\mathcal{H}_l^2\sim r_b \rho_b  V_b^4$ (which is valid for a penetration length  much smaller than the characteristic decay length of the evanescent wave, i.e., $L_p\ll r_b / \Lambda$ ), we find
\begin{align}
\mathcal{F}_r(r,\omega,l,m) &\sim \frac{\pi}{16} \frac{ \mathcal{N} \rho_b V_b ^4}{4 \pi r^2} \frac{b^4}{r_b^4} \frac{r_b}{N_0} \frac{e^{-\omega^2/4 \nu_p^2}}{\nu_p} \sqrt{l(l+1)}e^{-l(l+1) b^2/2 r_b^2}\nonumber \\
&\sim \frac{1}{4 \pi r^2}\frac{\mathcal{A} \mathcal{S}_p}{2} \frac{\rho_b V_b^3}{2} F_{R,l} \frac{e^{-\omega^2/4 \nu_p^2}}{\nu_p} e^{-l(l+1) b^2/2 r_b^2} \mbox{  ,}
\label{luminosity approximate}
\end{align} 
where $\mathcal{S}_p=\pi b^2$ is the area occupied by a single plume, $\rho_b V_b^3/2$ represents the plume kinetic energy flux and  $F_{R,l}=V_b k_{h,b}/ N_0$
is the Froude number at the top of the base of the convective zone, 
with $k_{h,b}=\sqrt{l(l+1)}/r_b$ the horizontal wavenumber. This latter represents the ratio of the advection by the mean plume flow to the buoyancy force at the top of the radiative zone when the plume penetrates into.
In the Sun, we find $F_{R,1}\sim 10^{-3}$ for $l=1$. 
In the following \sectionname{} \ref{application on a solar model}, we will demonstrate the validity of this expression (for a penetration length small enough) since the major part of the wave energy flux is distributed over angular degrees below the critical angular degree $l_{crit}$. Therefore, the implementation of angular momentum transport by plume-induced waves in a stellar evolution code will be easier using this simplified expression.

%
\section{The solar case}
%
\label{application on a solar model}

In this section, we use a solar model to compute the wave energy flux given by \eqref{spectral wave energy flux}. We analyze the effects of the plume characteristics on the excitation process and of the thermal transition length, $d$, on the wave transmission into the radiative zone (see \figurename{} \ref{schema situation}). 

\subsection{Input physics}
\label{input physics}
The calibrated solar model used here was obtained using the stellar evolution code CESTAM \citep{Marques2013}, with a mixing length parameter $\alpha_{MLT}=1.69$ and initial abundances $Y_0=0.25$ and $Z_0=0.013$. It provides the radial profile of all the quantities at the equilibrium that we need to compute the wave energy flux. The parameters computed from the model are the plume radius $b$, the plume velocity $V_b$ and the Brunt-Väisälä frequency $N_0$ at the bottom of the transition region $(r=r_d)$. Thus, using \eqref{largeur_panache} and \eqref{V_b} with $\rho_b \approx 150$ kg~m$^{-3}$ the density at the base of the convective zone and $r_b=5,1.10^{5}$ km the radius at the base of the convective zone, we find $b=10^4$ km and $V_b \approx 185 \mbox{ m~s}^{-1}$. In most of stellar codes, standard mixing-length treatment of the convection is local. It produces a very sharp thermal transition at the radiative/convective interface and so a very sharp slope for the Brunt-Väisälä frequency which remains difficult to estimate in this region. For this reason, we consider the average of the Brunt-Väisälä frequency below the base of the convective zone (as prescribed by the Schwarzschild's criterion) over a distance equal to $0.1 H_p$ in the radiative zone (see \sectionname{} \ref{sensitivity on input physics} for a discussion). Such a procedure leads to the value $N_0 = 282~\upmu{\rm H_z}$. To finish, there is still to choose three additional parameters, the plume filling factor $\mathcal{A}$, the penetration length $L_p$ and the characteristic plume turnover frequency $\nu_p$. For the filling factor, we use a reasonable value $\mathcal{A} \approx 0.1$ which corresponds to the values observed in numerical simulations of penetrative convection \citep[e.g.,][]{Brummell2002} and of the uppermost layers of the convective zone \citep{Stein1998}. This leads to a number of plumes equals to $\mathcal{N} \approx 1000$. We also take $L_p=0.05 H_p$, the upper limits given by \cite{Basu1997} for the Sun from seismic observations, and $\nu_p =1~\upmu{\rm H_z}$, a frequency around the convective turnover frequency (see discussion in \ref{plume lifetime}). 
%

\subsection{The case of a discontinuous transition $(d=0)$}
\label{result d=0}

\begin{figure}
\centering
\includegraphics[scale=0.48, trim= 0cm 0cm 0cm 1cm,clip]{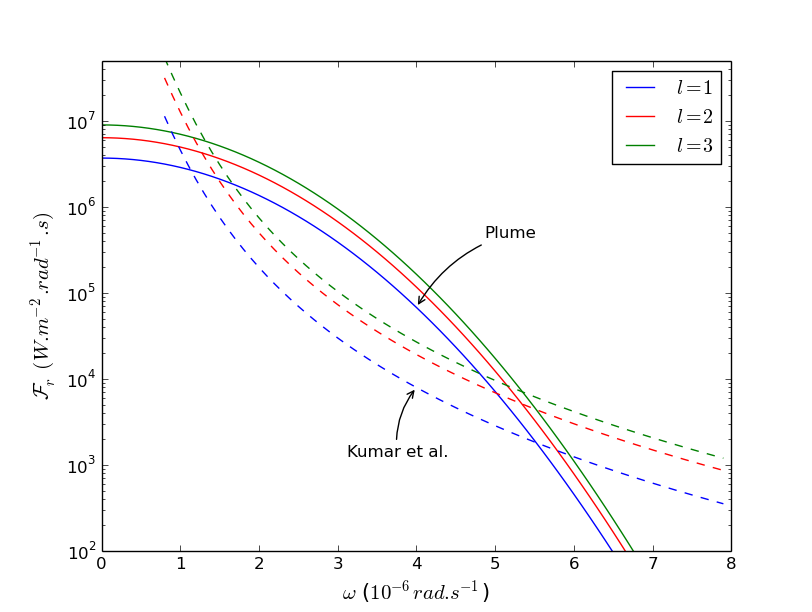} 
\caption{Mean radial wave energy flux per unit of frequency at the top of the radiative zone as a function of the radian frequency, for any azimuthal number $m$  and angular degrees $l=1,2$ and $3$, in the case of the plume-induced waves (solid lines) and the one of turbulence-induced waves from the formalism of \cite{Kumar1999} (dashed lines).}
\label{luminosity log}
\end{figure}

In a first approach, we set $d=0$. The result of the numerical calculation of \eqref{spectral wave energy flux} at the top of the radiative zone is plotted in \figurename{} \ref{luminosity log} (solid lines). The Gaussian profile of the spectrum comes from the term $\exp({-\omega^2}/4 \nu_p^2$). For $\omega \sim 10^{-6} \mbox{ rad~s}^{-1}$ and $l=1$, we read $\mathcal{F}_r \approx 3$ $10^{6}$~W~m$^{-2}$~rad$^{-1}$~s.
For comparison, using \eqref{luminosity approximate} and the value of the physical quantities given in \sectionname{} \ref{input physics}, we find $\mathcal{F}_r \approx 3.2$ $10^{6}$~W~m$^{-2}$~rad$^{-1}$~s.
As a check, we calculate the total wave energy flux integrated over a large frequency range, from $0$ to $25$ $10^{-6} \mbox{ rad~s}^{-1}$, for all the $(l,m)$ contributions, with $l\in [1,200]$ and $-l \le m \le l$. We find a low value amounting to about $1\%$ of the solar flux while the mean plume kinetic energy flux at the base of the convective zone, $\mathcal{A} \rho_b V_b^3/2$, represents about $40\%$ of the solar flux. These results justify a posteriori the assumption of no feedback from the waves on the plumes and verify the conservation of energy.

Hereafter, we specifically comment the effects of the different terms in \eqref{spectral wave energy flux} and the dependence of the spectrum on the plume characteristics $\nu_p$, $b$ and the penetration length $L_p$.
\begin{enumerate}
\begin{figure}
\centering
\includegraphics[scale=0.48, trim= 0cm 0cm 0cm 1cm,clip]{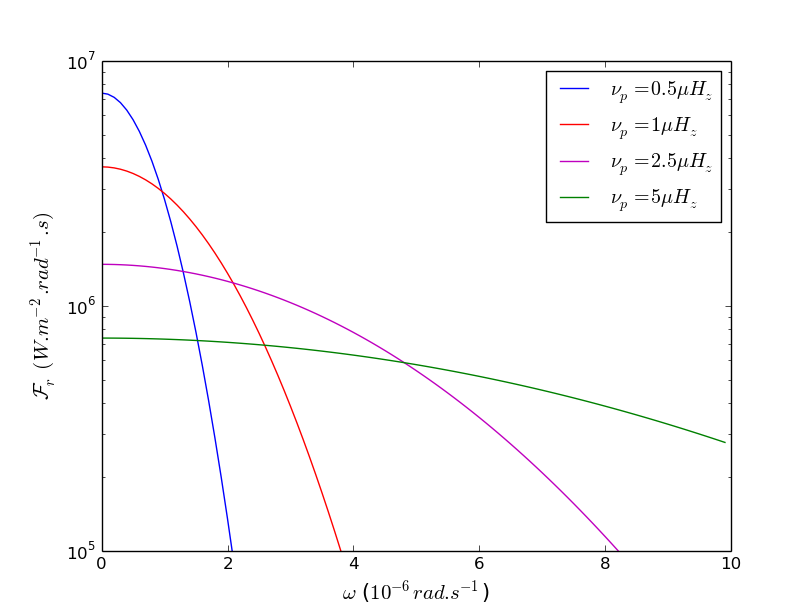} 
\caption{Mean radial wave energy flux per unit of frequency at the top of the radiative zone as a function of the radian frequency, for any azimuthal number $m$, an angular degree $l=1$ and varying plume frequencies.}
\label{lum vs nu_p}
\end{figure}
\item {\it{Time correlation (\figurename{} \ref{lum vs nu_p})}:} 	We consider four values in range of one order of magnitude around $1$~$\upmu{\rm H_z}$ for the plume turnover frequency, $\nu_p=0.5, 1, 2.5$ and
$5$~$\upmu{\rm H_z}$. The result is plotted in \figurename{} \ref{lum vs nu_p} as a function of the cyclic frequency. As said above, an increase of $\nu_p$ involves a redistribution of the wave energy over higher frequencies at the expense of the lower ones since the total wave energy does not depend on the plume lifetime (under the assumption of stationarity). Its influence on the shape of the wave energy flux is significative: an increase by a factor five transforms a bell-shaped spectrum into an nearly flat one in the considered frequency range. 

\begin{figure}
\centering
\includegraphics[scale=0.48, trim= 0cm 0cm 0cm 1cm]{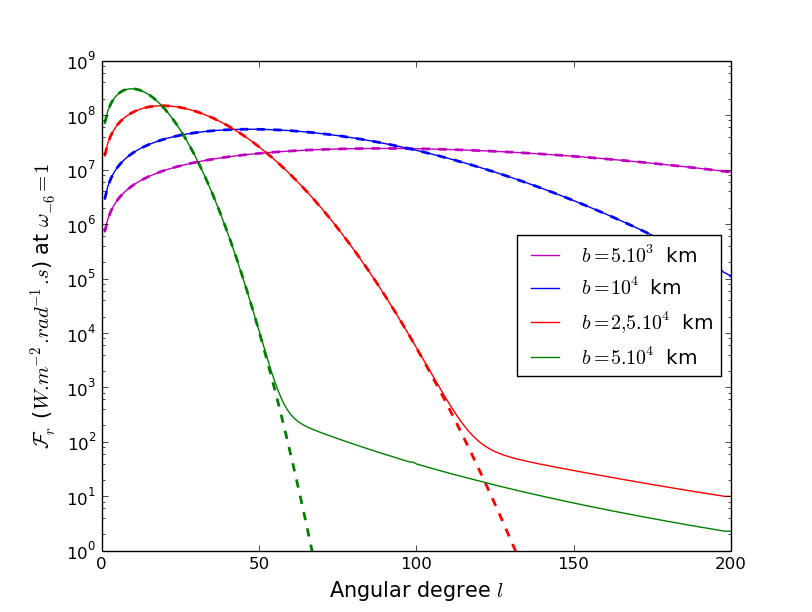} 
\caption{Mean radial wave energy flux at the top of the radiative zone and at the frequency $\omega=10^{-6} \mbox{ rad~s}^{-1}$, as a function of the angular degree and for varying plume radii, $b$. The numerical results from \eqref{horizontal average} (solid lines) are compared to the results using \eqref{B_l theorique} (dashed lines). The filling factor is supposed to be constant so that a change in the plume radius results in a change of the number of plumes.}
\label{horizontal correlation}
\end{figure}
\item {\it{Horizontal correlation (\figurename{} \ref{horizontal correlation})}:} We assume a constant filling factor, $\mathcal{A} \approx 0.1$, so that a change in the plume radius involves a change in the number of plumes. We consider four values in a range of one order of magnitude around $10^4$ km for the plume radius, $b=0.5,1,2.5$ and $5$ $10^4 $ km, leading to a number of plumes $\mathcal{N}=4000,1000, 160$ and $40$, respectively. Although the last value seems unrealistic for the Sun, we use it to study the general behavior of the wave energy spectrum with respect to the plume radius. The wave energy flux is maximal for a degree that corresponds to the horizontal extension of the plume, $l_{max}=~r_b/b$, independently of the frequency. We find $l_{max}=~102,51,20$ and $10$, respectively. that agrees with the numerical results plotted in \figurename{} \ref{horizontal correlation} as a function of the angular degree at the fixed frequency of $10^{-6} \mbox{ rad~s}^{-1}$. As mentioned in \sectionname{} \ref{wave energy flux main}, we can verify on the same figure that the use of \eqref{B_l theorique} (dashed lines)  instead of \eqref{horizontal average} (solid lines) for the term $\mathcal{B}_l$ is valid below a critical degree $l_{crit}=2 \pi r_b /b \approx 125$ and $63$ for $b=2.5$ and $5$ $10^4 $ km, respectively. Furthermore, a decrease in the plume radius induces an increase of the wave energy flux at low degrees at the expense of higher degrees. The total wave energy flux integrated over the frequency range $\omega=0-25$~$10^{-6} \mbox{ rad~s}^{-1}$ and for the degrees $l=1-200$ is equal to $1.3 \%, 1 \%, 0.5 \%$ and $0.25 \%$ of the solar flux for $b=0.5,1,2.5$ and $5$ $10^4$ km, respectively. Thus, the total wave energy flux decreases with larger plume radii, particularly because the number of penetrating plumes decreases in order to keep a constant filling factor. 
\begin{figure}
\centering
\includegraphics[scale=0.48, trim= 0cm 0cm 0cm 1cm,clip]{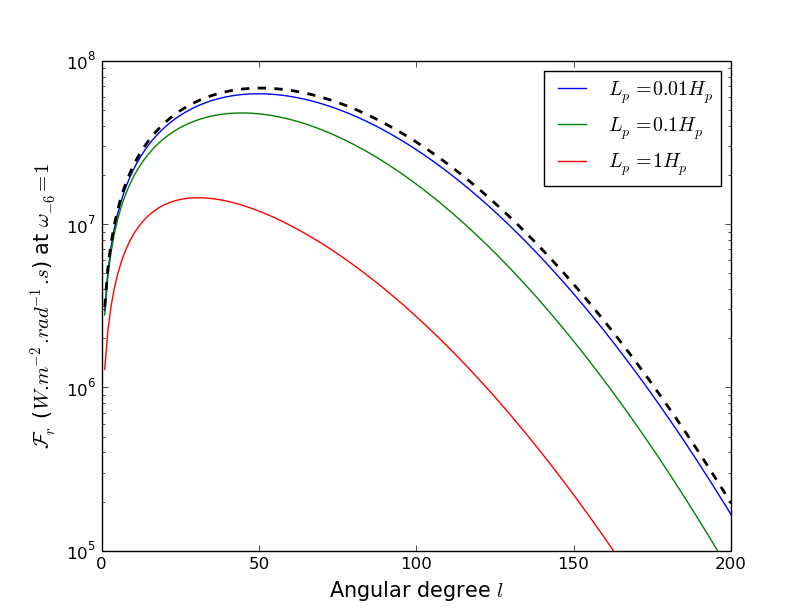} 
\caption{Mean radial wave energy flux at the frequency $\omega=10^{-6} \mbox{ rad~s}^{-1}$, as a function of the angular degree $l$ for different penetration lengths. The black dashed line represents the result obtained using \eqref{luminosity approximate}.}
\label{H_l vs l}
\end{figure}
\item {\it{Driving term (\figurename{} \ref{H_l vs l})}:} We choose three values for the penetration length within a range of two orders of magnitude, $L_p=0.01,0.1$ and $1H_p$. The wave energy spectrum depends on this parameter only through the term $H_l$ in \eqref{H_l radial}. The result is plotted in \figurename{} \ref{H_l vs l} as a function of the angular degree at the fixed frequency of $10^{-6} \mbox{ rad~s}^{-1}$. An increase of the penetration length
induces a decrease of the transmission of power from the plume into the waves. The higher the angular degree, the higher the decrease. Although this effect is moderate for low degrees (decrease by a factor about two for $L_p= H_p$ and $l<50$), it becomes critical for higher degrees (decrease of about one order of magnitude for $L_p= H_p$ and $l>150$). Moreover, this effect results in a shift of the maximum of the wave energy spectrum to lower angular degrees with an increasing penetration length. It goes from $l_{max}\approx 50$ for $L_p=0.01 H_p$ to $l_{max} \approx 30$ for $L_p=1 H_p$. The total wave energy flux integrated over the frequency range $\omega= 0 - 25$~$10^{-6} \mbox{ rad~s}^{-1}$ for $l=1-200$ represents about $1.3 \%, 0.8 \%$ and $0.2 \%$ of the solar flux for $L_p=0.01,0.1$ and $1H_p$, respectively. Finally, we check that the assumption made on $\mathcal{H}_l$ to derive \eqref{luminosity approximate} (black dashed line) is valid if $L_p\ll r_b/ \Lambda$, i.e., $l \ll r_b / L_p = 940, 94$ and $9$ for the three considered cases.
\end{enumerate}

In summary, \eqref{luminosity approximate} has been shown to be a good approximation of \eqref{spectral wave energy flux} provided that $l\ll l_{crit}$ and $L_p \ll r_b /\Lambda$. We have seen that $\nu_p$, $b$ and $L_p$ have a significative effect on the shape of the wave energy spectrum and the total energy transferred into the waves. This could have serious consequences especially for the issue of transport by IGW. Last, the set of values used hereabove for $\nu_p$, $b$ and $L_p$ will enable us to compare penetrative convection to driving by turbulent convection considering a wide domain for these parameters. This will be discussed in \sectionname{} \ref{discussion}.

\subsection{Effect of a thermal transition layer $(d \ne 0)$}
\label{result d ne 0}
\begin{figure}
\centering
\includegraphics[scale=0.48, trim= 0cm 0cm 0cm 1cm,clip]{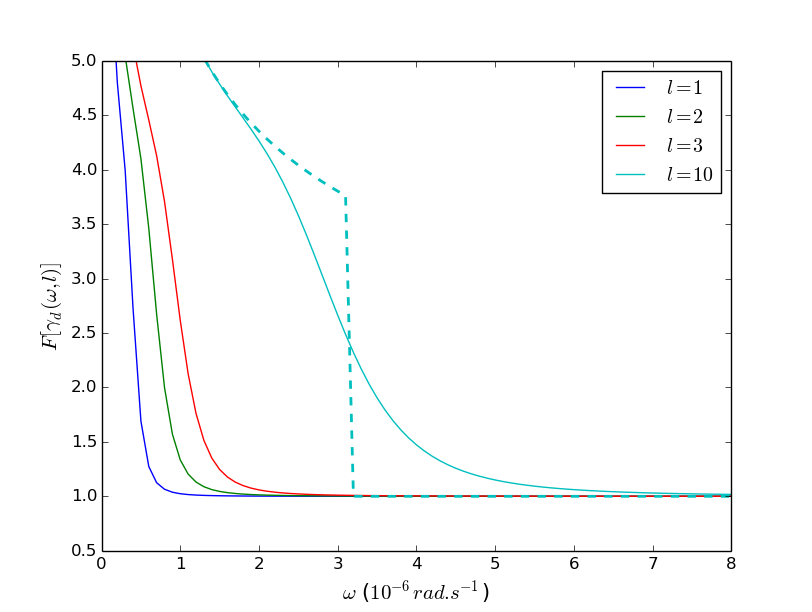} 
\caption{Transmission function defined in \eqref{smooth transition function} for $d=0.01 H_p$ (solid line), as a function of the radian frequency for $l=1,2,3$ and $10$. It is compared to the result obtained using \eqref{transition function} for $l=10$ (dashed line).}
\label{ratio smooth sharp}
\end{figure}
We now consider the case of a thermal transition at the radiative/convective interface ($d \ne0$ in \figurename{} \ref{schema situation}). We choose a transition lentgh $d=500$ km, i.e., $d \approx 0.01 H_p= L_p/5$ (see \sectionname{} \ref{sensitivity on input physics} for a discussion).

In the derivation of \eqref{transition function}, we have considered the limiting cases of very sharp transition and of a very smooth one; the first one, corresponding to $\gamma_d(\omega,l) \ll 1$ in \eqref{gamma_d}, allows for using a first-order Taylor expansion of Airy functions at $0$ in the transition layer, while the second one, corresponding to $\gamma_d(\omega,l) \gg 1$, enables us to use their asymptotic expansion at $-\infty$ (see \appendixname{} \ref{continuity of the wave function} for details). To avoid a discontinuity between the two regimes at $\gamma_d(\omega,l)=1$ (as in \eqref{transition function}, see the dashed line in \figurename{} \ref{ratio smooth sharp}) and to be more realistic, we assume that the switchover from the first case to the second one is realized on the range $\gamma_d(\omega,l)\in [0.80,1.20]$. Instead of using \eqref{transition function},  we define the transmission function, $F$, as a weighted average of the two limiting cases,
\begin{equation}
F \left(\gamma_d \right)=\left[ 1 - g(\gamma_d)\right]+g(\gamma_d)\tilde{f}(\gamma_d)\mbox{  ,}
\label{smooth transition function}
\end{equation}
where
\begin{align}
g(x)&=\left\{\tanh\left[(x-1)/0.20\right]+1\right\}/2\\
\tilde{f}(x)&=D \sqrt{x} \left(1-\omega^2/N_0^2\right)^{-1} \mbox{  ,}
\end{align}
with $g$ the weighting function.
The result is plotted in \figurename{} \ref{ratio smooth sharp} as a function of the cyclic frequency for $l=1, 2, 3$ and $10$. Physically, it represents the ratio of the wave energy flux obtained in the case of a linear transition $(d= 500$km$)$, to the flux in the case of a discontinuous transiton $(d=0)$. We see that the wave energy flux can be enhanced by factor two to five, depending on the ratio $k_h d N_0/\omega$. Indeed, the higher is this ratio, the smaller is the radial wavelength (relatively to $d$), and the smoother is the transition from the point of view of this component $(\omega,l)$. This result agrees with the work of \cite{Lecoanet2013} who took into account a smooth thermal transition in the case of wave driving by turbulent pressure.

The total wave energy flux at the top of the radiative zone integrated over the frequency range $\omega= 2-25$ $10^{-6} \mbox{ rad~s}^{-1}$ for $l=1-200$ represents about $1.2\%$ of the solar flux, which is about $10$ times higher than in the case of a discontinuous transition. For $\omega=0-25$ $10^{-6} \mbox{ rad~s}^{-1}$ and $l=1-200$, integration gives about $11\%$ of the solar flux, i.e., about one fourth of the mean plume kinetic energy flux at the base of the convective zone. However, the latter result is likely to be overestimated because of unmodeled physical processes at very low frequencies (see \sectionname{} \ref{limits of the model} for a discussion).

Hence, a smooth thermal transition at the radiative/convective interface is able to significatively enhance the transmission of the wave energy flux into the radiative zone. It also has an impact on the shape of the wave energy spectrum at the top of the radiative zone since it does not affect in the same way the different wave components. Therefore, the length of the thermal adjustment layer, $d$, is a key parameter in the generation process of IGW. Unfortunately, numerical simulations of penetrative convective are not currently able to quantify the extent of this thermal adjustment layer in the solar parameters regime \citep[e.g.,][]{Brummell2002}. Nevertheless, several semianalytical models of penetrative convection can give us some hints to its value (see \sectionname{} \ref{sensitivity on input physics}).

%
\section{Discussion}
%
\label{discussion}

In this section, we compare our results with those based on the excitation by turbulent pressure through the whole convective zone as formulated by \cite{Kumar1999}. We also compare them with the studies of \cite{Press1981} and \cite{Lecoanet2013} that focus on the excitation by convective eddies at the radiative/convective boundary.
Lastly, we investigate the sensitivity of the present excitation model to the input physics and discuss its limits.

\subsection{Comparison with turbulence-induced wave energy flux \citep{Kumar1999}.}
\label{kumar turbulence}

\begin{figure}
\centering
\includegraphics[scale=0.48, trim= 0cm 0cm 0cm 1cm,clip]{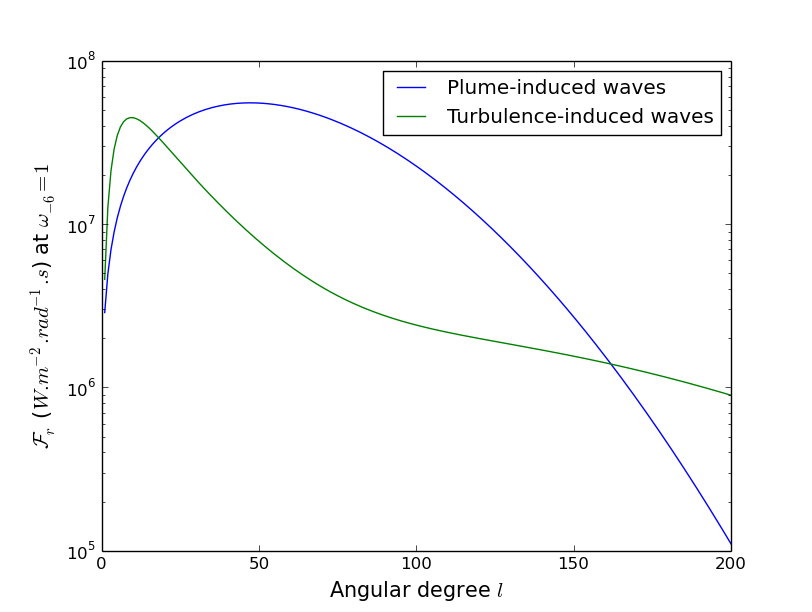} 
\caption{Mean radial wave energy flux per unit of frequency, as a function of the angular degree $l$ at the frequency $\omega=10^{-6} \mbox{ rad~s}^{-1}$, generated by penetrative convection (in blue) and by turbulent pressure in the convective zone (in green) from the formalism of \cite{Kumar1999}. }
\label{luminosity vs l}
\end{figure}

IGW can also be generated by turbulent pressure in the convective zone. \cite{Kumar1999} proposed a model of excitation by Reynold's stress due to turbulent motions in the convective zone. The energy cascade is supposed to follow the Kolmogorov's spectrum. The wave energy flux per unit of frequency at the base of the convective zone for an angular degree $l$ and an azimuthal number $m$ is given by
\begin{align}
\mathcal{F}_{E}^c(\omega,l,m)=&\frac{\omega^2}{4\pi} \int_{r_b}^{R_{s}}
 \mathrm{d} r\frac{ \rho^2}{r^2}
 \left[\left( \frac{\partial \xi_r}{\partial r}\right)^2+l(l+1)\left( \frac{\partial \xi_h}{\partial r}\right)^2\right]\nonumber\\
&\times\exp\left[-\frac{h_{\omega}^2 l(l+1)}{2 r^2}\right] \frac{\mathrm{v}^3 L^4}{1+(\omega \tau_L)^{15/2}} \mbox{  ,}
\label{flux Kumar}
\end{align}
where $\xi_r$ and $\xi_h$ are the radial and horizontal wave displacements normalized to unit IGW energy flux just below the convective zone, $R_s$ is the outer radius of the convective zone, $\mathrm{v}$ is the velocity of the biggest convective eddies with a size $L$ and a turnover time $\tau_L=L/\mathrm{v}$, and $h_{\omega}=L\min[1,2(\omega\tau_L)^{-3/2}]$ is the size of the convective eddies with frequency $\omega$ at radius $r$. The lower limit frequency for these waves is the convective turnover frequency at the base of the convective zone, $\omega_c$,  such as $\mathcal{F}_{E}^c(\omega < \omega_c,l,m) = 0$. To compute this expression, $L$ and $\mathrm{v}$ are deduced from the MLT and the wave displacement is approximated by the normalized WKB wave functions as in Eq. (25) of \cite{Kumar1999}. In our solar model, we find a convective turnover frequency $\omega_c \approx 0.8$ $10^{-6} \mbox{ rad~s}^{-1}$ and a convective velocity $\mathrm{v}_c \approx 25$~m~s${}^{-1}$ at the base of the convective zone, which corresponds to a convective flux equal to about $5\%$ of the solar flux. The result for the solar model is plotted in \figurename{} \ref{luminosity log} (dashed lines) as a function of the cyclic frequency for $l=1,2$ and $3$, and in \figurename{} \ref{luminosity vs l} as a function of the angular degree at $\omega=10^{-6} \mbox{ rad~s}^{-1}$, since we consider that most of the wave energy is distributed around this frequency in both cases. 

The total turbulence-induced wave energy flux integrated over $\omega=0.8-25$ $10^{-6} \mbox{ rad~s}^{-1}$ for angular degrees $l=1-200$ is equal to $0.1\%$ of the solar flux. With $0.6\%$ of the solar flux, the excitation by penetrative convection is 6 times more efficient in this frequency range. By integrating over the frequency range $1-25$ $10^{-6} \mbox{ rad~s}^{-1}$, the total turbulence-induced wave energy flux falls to $0.02 \%$ of the solar flux, mainly because most of the energy is contained at low frequencies and because the spectrum decreases rapidly around $10^{-6} \mbox{ rad~s}^{-1}$. In turn, the plume-induced wave energy flux remains at $0.5\%$ of the solar flux. Indeed, the turbulence-induced wave energy flux is about one order of magnitude smaller than the plume-induced one in the range $2-5$ $10^{-6} \mbox{ rad~s}^{-1}$ (\figurename{} \ref{luminosity log}), but it dominates otherwise. In \figurename{} \ref{luminosity vs l}, we compare the wave energy flux for both driving processes as a function of the horizontal degree $l$. In the case of the turbulence-induced waves, the wave energy flux is maximal for $l\sim5$ which corresponds to the size of the biggest convective eddies as given by the MLT, $L = \alpha_{MLT} H_p \sim 10^5$ km, while the plume-induced wave energy flux peaks at $l\sim50$ because of the smaller horizontal extent of the plumes.

In \figurename{} \ref{luminosity log}, the plume-induced wave energy flux drops drastically below the turbulence-induced one for frequencies higher than $5$ $10^{-6} \mbox{ rad~s}^{-1}$. In a similar way, in \figurename{} \ref{luminosity vs l}, the turbulence-induced wave energy flux overtakes the plume-induced one for degrees higher than a few characteristic widths of the plume-induced wave spectrum, $r_b/b$. In reality, the flow of the plume should become turbulent on higher temporal (and spatial) frequencies and should generate waves by Reynold's stress due to shear flow, as proposed by \cite{Montalban2000}. Indeed, we assumed a coherent plume flow, with a spatial and a temporal characteristic extent, and neglect the turbulence inside the plume. As a consequence, the present model does not take into account plume-related perturbations in the high frequency domain. The power transmitted from the plume into the waves on temporal (spatial) frequencies higher than few $\nu_p$ (few $r_b/b$) vanishes rapidly and the excitation by convective turbulence then can become predominant.

Finally, wave driving by penetrative convection is five to ten times more efficient than the excitation by turbulence as formulated by \cite{Kumar1999}. While turbulence excites preferentially at very low frequencies, penetrative convection efficiently produces waves on a larger frequency range (it nevertheless depends on the value of $\nu_p$ relatively to $\omega_c$). The plume-induced wave spectrum peaks at higher angular degrees than the turbulence-induced one because  the plume radius is smaller than the size of the energy-bearing convective eddies. We emphasize that the same conclusions are reached with the wide set of values considered for $\nu_p$, $b$ and $L_p$ in \sectionname{} \ref{result d=0}. All the differences observed between wave energy flux spectrum produced by both mechanisms will have consequences on the transport of angular momentum by IGW which are discussed in the following section. 

%

\subsection{Comparison with models of excitation by turbulent eddies at the radiative/convective boundary}

The excitation model as proposed by \cite{Kumar1999} assumes that waves are generated by Reynold's stress in
each layer of the whole convective region and tunnel towards the radiative zone where they can propagate. While this model is used in most of the actual computation of angular momentum transport by internal gravity waves, it is also interesting to compare our results with the excitation models of \cite{Press1981} and \cite{Lecoanet2013}. Indeed, their approaches are similar to the one used in the present model in the sense that they focus on the driving of waves in the confined overshoot region.

\cite{Press1981} considered the convective motions of the biggest eddies above the boundary as the driving mechanism. The eddies are characterized by a frequency $\omega_c$, a size $\lambda=2 \pi/k_c$ and a velocity $\mathrm{v}_c=\omega_c/k_c$. He assumed that the eddies generate waves with a frequency $\omega \sim \omega_c$ and a horizontal wavenumber $k_h \sim k_c$, and he matched the turbulent pressure with the wave  pressure perturbation, i.e., $p^\prime~\sim~\rho~ \mathrm{v}_c^2~\sim~\rho~ (\omega/k_h)^2$. Assuming a discontinuous thermal transition ($d=0$ in \figurename{}~\ref{schema situation}) and using the continuity of the Eulerian perturbation of pressure at the boundary, the amplitude of the vertical wave displacement just below the radiative/convective interface is deduced from the relation $\xi_r=p^\prime / Z_v$, where $Z_v=\rho N_0 \omega / k_h$ is the wave impedance at the top of the radiative zone. 
Using the Boussinesq's approximation, the specific wave energy just below the boundary equals $\mathcal{E} \sim \rho N_0^2 \xi_r^2$. The radial wave energy flux is then obtained by multiplying $\mathcal{E}$ with the vertical group velocity given by \eqref{group_velocity} so that we obtain
\begin{align}
\mathcal{F}_r^{P81}\sim \rho \frac{\omega^3}{ k_h^3} \frac{\omega}{ (N_0^2-\omega^2)^{1/2}} \sim \rho \mathrm{v}_c^3\frac{ \mathrm{v_c} k_h }{(N_0^2-\omega^2)^{1/2}} \mbox{  .} \label{flux press}
\end{align}
Equation \ref{flux press} have some similarities and differences with our model and its simplified expression \eqref{luminosity approximate}:
\begin{enumerate}
\item Equation \ref{flux press} is equal to the product of the mechanical convective flux with the Froude number (or the convective Mach number) at the base of the convective zone, which also plays the role of a transmission factor. This is similar to the product of the plume kinetic energy flux $\rho V_b^3$ with $F_{R,l}$ in \eqref{luminosity approximate} with the substitution $\mathrm{v}_c \leftarrow V_b$.

\item Equation \ref{flux press} contains no equivalent of the exponential terms in \eqref{luminosity approximate} since \cite{Press1981} assumed that waves are efficiently excited in a narrow frequency range around $\omega_c$.

\item Excitation by convective eddies occurs in each element of surface of the spherical shell at the radiative/convective boundary whereas convective penetration generates waves only in $\mathcal{N}$ localized regions of the spherical shell. As a result, \eqref{luminosity approximate} differs from \eqref{flux press} by an additional geometrical factor, $\mathcal{A} S_{p} / 4 \pi r^2$, which represents the fraction of the area occupied by the ensemble of plumes.
\end{enumerate}
The excitation models proposed by \cite{Garcia1991} and \cite{Zahn1997} are very similar to the one of \cite{Press1981}, except that they consider in addition a Kolmogorov's distribution of convective eddies with an incoherent behavior. 

More recently, \cite{Lecoanet2013} investigated also the excitation of internal gravity waves by turbulent convection. In their model, the wave amplitude is derived by considering the turbulent Reynold's stress as the driving mechanism in a similar way to \cite{Kumar1999}. However, their approach differs from the latter since they focused on the excitation of waves in the overshooting layer and took into acount the thermal transition layer at the radiative/convective interface (see \figurename{} \ref{schema situation}). 
Moreover, they assumed that the injection length scale of the turbulent cascade at the boundary is of the same order of magnitude as the thickness of the overshooting region (i.e., smaller than the usual mixing length of the order of $H_p$) while conserving the same convective velocity as given by the MLT. In other words, the velocity of the eddies with a typical size $h < H_p$ at the boundary is higher than in the case of the description of the turbulence by the MLT as used in \cite{Kumar1999}. The spectral density of the wave energy flux in the case of a smooth linear thermal transition, Eq. (42) in \cite{Lecoanet2013}, is given by
\begin{align}
\mathcal{F}_r^{L13}(\omega,l) &\sim\rho \mathrm{v}_c^3 \left(\frac{\omega_c}{N_0}\right)^{2/3} (k_h H^{\star}_{l,\omega})^4 \left(\frac{\omega}{\omega_c} \right)^{-41/6} \frac{1}{\omega} \left(k_h d\right)^{1/3} \mbox{  ,}
\end{align}
which we can rewrite as
\begin{align}
\mathcal{F}_r^{L13}(\omega,l)&\sim \rho \mathrm{v}_c^3 \frac{ \mathrm{v}_c k_c}{N_0} \left( \frac{k_h}{k_c} \right)^{17/3}\left( \frac{H^{\star}_{l,\omega}}{H_p} \right)^4 \left(\frac{\omega}{\omega_c} \right)^{-41/6} \frac{1}{\omega}\left(\frac{k_c N_0 d}{\omega_c}\right)^{1/3} \mbox{,} \label{flux lecoanet}
\end{align}
with $\omega \ge \omega_c$ and $k_h \le k_{h,max}(\omega)$, where we have used $ \mathrm{v}_c =~\omega_c / k_c$ and $k_c \sim 1/H_p$, the typical convective scales as given by the MLT. We have also defined $H^{\star}_{l,\omega}$ which depends on the characteristic length scales of the turbulence that is considered in the excitation zone ($H^{\star}_{l,\omega}= \alpha_{MLT} H_p$ in the case of the MLT). \cite{Lecoanet2013} showed that such considerations may enhance the wave flux by a factor two to five. The excitation model of \cite{Lecoanet2013} differs from the present one by several points: 
\begin{enumerate}
\setcounter{enumi}{3}
\item The  points 1 and 3 mentioned hereabove are still valid in this case.
\item  Equation \ref{flux lecoanet} depends on $\omega$ and $k_h$ as power laws coming from the assumption of a Kolmogorov's spectrum for the turbulence, whereas \eqref{luminosity approximate} depends on them as exponential laws coming from the Gaussian profile used to model the plume velocity in the penetration region.
\item  The last term in \eqref{flux lecoanet} is related to the effect of the thermal adjustment layer of length $d$ on the wave transmission into the radiative zone. As a consequence, \eqref{flux lecoanet} is proportional to $(d/N_0^2)^{1/3}$, which is consistent with our model.
\end{enumerate}

Finally, the expressions derived by \cite{Press1981} and \cite{Lecoanet2013} for the wave energy flux at the top of the radiative zone scale as $\mathrm{v}_c^4$ whereas the plume-induced wave energy flux depends on the plume velocity as $V_b^4$. With a plume velocity one order of magnitude higher than the convective velocity in the solar case, we can conclude that the driving by penetrative plumes is more efficient than the excitation by turbulent eddies at the radiative/convective boundary as in the model proposed by \cite{Press1981} and \cite{Lecoanet2013}.

\subsection{Efficiency of the angular momentum transport by plume-induced IGW}

%
\begin{figure}
\centering
\includegraphics[scale=0.48, trim= 0cm 0cm 0cm 1cm,clip]{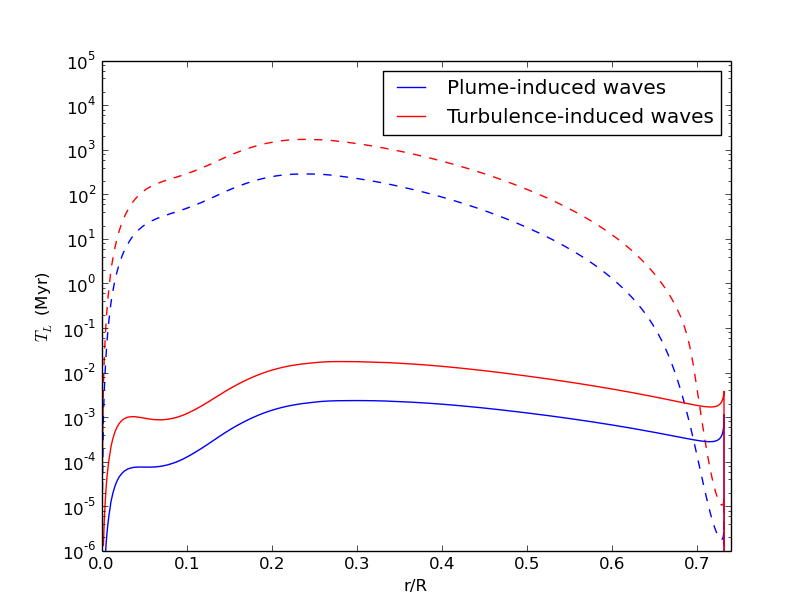} 
\caption{Characteristic time scale of evolution of the rotation rate in the case of a low differential rotation $\delta \Omega=0.15~10^{-6} \mbox{ rad~s}^{-1}$ (dashed lines) and a stronger one  $\delta \Omega=10^{-6} \mbox{ rad~s}^{-1}$ (solid lines). We compare the results obtained with a plume-induced wave spectrum (blue) and with a turbulence-induced wave spectrum (red) from the formalism of \cite{Kumar1999}.
 }
\label{timescale diff low}
\end{figure}

IGW are able to carry angular momentum in the radiative zone and to modify the mean rotation rate. The wave flux of angular momentum is deduced from the wave energy flux through the relation \citep{Zahn1997}
\begin{equation}
\mathcal{F}_J(r,\omega,l,m)=\frac{m}{\omega} \mathcal{F}_E(r,\omega,l,m) \mbox{  .}
\end{equation}
In the adiabatic case, the wave luminosity of angular momentum is conserved through the whole star and no exchange is possible with the mean flow. However, as they propagate towards the center of the star, IGW are radiatively damped from the top of the radiative zone. The total wave flux of angular momentum is deduced at each depth from the one at the top of the radiative zone, modulated by a damping term
\begin{equation}
\mathcal{F}_J^T(r)=\sum_{l,m} \int \mathcal{F}_J(r_d,\omega,l,m) e^{-\tau(r,\hat{\omega},l)}\mathrm{d} \omega \mbox{  .}
\end{equation}
\cite{Press1981} investigated the damping of IGW and found in the quasi-adiabatic approximation
\begin{equation}
\tau(r,\hat{\omega},l)=\left[l(l+1)\right]^{3/2}\int_{r}^{r_d} K  \frac{N^3}{\hat{\omega}^4}\left(\frac{N^2}{N^2-\hat{\omega}^2}\right)^{1/2} \frac{\mathrm{d} r}{r^3} \mbox{  ,}
\end{equation}
where $K$ is the radiative diffusion coefficient and
\begin{equation}
\hat{\omega}(r,\omega,m)=\omega-m\delta \Omega(r)
\end{equation}
is the Doppler-shifted frequency in the corotating frame, with $\delta \Omega$ the differential rotation with respect to rotation rate at the base of the convective zone. Hence, IGW can deposit angular momentum in presence of differential rotation since prograde waves ($m>0$) and retrograde waves ($m<0$) are asymmetrically damped.

The transport of angular momentum follows an advective-diffusive equation \citep[e.g.,][]{Maeder2009}. Solving numerically the problem is out of scope here but a preliminary study can give some clues on the efficiency of the transport by waves. The characteristic time scale, $T_L$, on which IGW can affect the rotation mean-flow can be estimated through \citep[e.g.,][]{Belkacem2015b}
\begin{equation}
T_L(r)\sim\frac{\rho r^2 \delta \Omega}{\dot{J}} \mbox{  ,}
\label{rotation time scale}
\end{equation}
where $\dot{J}$ represents the divergence of the total mean radial wave flux of angular momentum
\begin{equation}
\dot{J}=\frac{1}{r^2} \frac{\partial}{\partial r} \left( r^2 \mathcal{F}_J^T(r) \right) \mbox{  .}
\end{equation}
In order to get an order of magnitude and to simplify the calculation, we assume that the radiative zone rotates like a solid body, so that $\delta \Omega$ is constant in the whole radiative zone. We consider the case of a low differential rotation with respect to the base of the convective zone, $\delta \Omega=0.15~10^{-6} \mbox{ rad~s}^{-1}$ (hereafter, case {\it l}), and the case of a stronger differential rotation, $\delta \Omega=~10^{-6} \mbox{ rad~s}^{-1}$ (hereafter, case {\it s}). The case {\it l} is almost similar to the initial rotation profile in \citet[Fig. 3]{Talon2002}. We assume also that the waves deposit all their angular momentum when they meet a critical layer (i.e., $\hat{\omega}=0$) and that they only travel one-way towards the core of the star. We take into account only degrees from $1$ to $50$ since we consider it is representative of the wave spectrum in both processes of excitation. The frequency range of integration is taken between $1$ and $6$ $10^{-6} \mbox{ rad~s}^{-1}$, since most of the wave energy is concentrated in this range and since theory is uncertain below. The result is plotted in \figurename{} \ref{timescale diff low}.

We find that transport by plume-induced waves is able to considerably affect the internal rotation rate in the Sun in less than 0.1 Gyr in the case~{\it l} and in less than 0.01 Myr in the case~{\it s}. Just below the convective envelope, characteristic time scale is of the order of the year; this is the result of the rapid damping of very high degree and low frequency prograde waves; this can generate a shear layer oscillation as obtained, for instance, in \cite{Talon2002}. In the radiative interior, plume-induced waves can affect the rotation mean-flow on time scales about one order of magnitude shorter than turbulence-induced ones. Note that the core could be decelerated in less than 0.01 Gyr in the case {\it l} and in less than 1000 yrs in the case {\it s} whatever the excitation process is considered because the specific angular momentum becomes weaker and weaker in the innermost layers as it decreases as $r^2$. Moreover, the comparison between both cases {\it l} and {\it s} emphasizes the character of IGW which tends to work against strong rotation rate gradients. The stronger the differential rotation, the shorter the time scale on which IGW can affect the rotation profile.

Therefore, we conclude that penetrative convection can generate waves able to affect the rotation profile of solar-like stars over time scales much shorter than their lifetime on the main sequence (around 10 Gyr). Moreover, transport of angular momentum by plume-induced waves is more efficient than by turbulence-induced ones. A more quantitative study including all mechanisms of transport, as meridional circulation and shear-induced turbulence, have to be done to confirm this result.

\subsection{Sensitivity on input physics} 
\label{sensitivity on input physics}

The proposed wave excitation model depends on several physical quantities, like the plume characteristics $b$, $\nu_p$, $V_b$, and on other parameters such as $N_0$, $\mathcal{A}$, $L_p$ and $d$. We discuss here the sensitivity of the model to these quantities.

The effect of $b$ and $\nu_p$ has already been investigated in \sectionname{} \ref{application on a solar model}. We have shown that the total wave energy flux does not depend on the value of $\nu_p$ since we assume a stationary process of excitation and that it is almost inversely proportional with $b$. However, they have a real impact on the shape of the spectrum. From \eqref{luminosity approximate}, the smaller $b$ ($\nu_p$), the larger (the smaller) the width of the spectrum in the spatial (temporal) frequency domain. Moreover, the amplitude of the spectrum depends on $1/\nu_p$ and $b^2$. Multiplication by a factor 2 for $\nu_p$ and $b$ leads to a multiplication by a factor 0.5 and 4, respectively. Thus, the error made on the values of $\nu_p$ and $b$ is not negligible in the estimate of the wave energy spectrum.
The plume velocity $V_b$ at the entry of the penetration zone is another important plume characteristic which highly influences the wave energy flux. Indeed, following \eqref{luminosity approximate}, the wave energy flux is proportional to $V_b^4$. This means that an increase of 10\%, 20\% and 30\% in $V_b$ leads to an increase of about 50\%, 250\% and 500\% in the wave energy flux, respectively. Similarly, a decrease of 10\%, 20\% and 30\% for $V_b$ leads to a decline of about 35\%, 60\% and 75\%. We think the model of turbulent plumes of \citep{Rieutord1995} is reliable enough to give reasonable values for $V_b$ and $b$ which will not affect too much the wave energy flux, i.e., by a factor between 0.25 and 4. Note that the same issue arises in the excitation by turbulence since the wave power spectrum depends on the convective velocity to the power 3 and on the size of the energy-bearing to the power 4 in \eqref{flux Kumar}. Uncertainties linked to their determination by the MLT have also a serious effect on the estimate of the wave energy flux in their model. 

Among the free parameters, the filling factor represents the fraction of the area occupied by the ensemble of plumes at the base of the convective zone. In our model, the wave energy flux is proportional to $\mathcal{A}$. Conservation of matter imposes $\mathcal{A} < 0.5$ since it exists upflows less dense and slower than the plume downflows. We use a reasonable value of about $\mathcal{A} \approx 0.1$ and expect it not to vary too much around this value given what is observed in numerical simulations of the uppermost layers of the convective zone \citep{Stein1998}. Given \sectionname{} \ref{kumar turbulence}, a filling factor lower than $0.01$ is required for penetrative convection to become negligible compared to the excitation by turbulent convection. 

The penetration length, $L_p$ (see \figurename{} \ref{schema situation}), has been theoretically shown to be of the order of magnitude of $H_p$ by \cite{Zahn1991}. \cite{Basu1997} observationally found a lower limit equals to $0.05 H_p$ in the Sun, i.e., more than one order of magnitude below the theoretical prediction. The wide range of values for $L_p$ considered in \sectionname{} \ref{result d=0} reflects this discrepancy between theory and observations. From \figurename{} \ref{H_l vs l}, we see that a change in $L_p$ does not influence too much the wave energy spectrum for low angular degrees. A multiplication by 1000 of $L_p$ leads to a division by about five of the wave energy flux for $l<50$. However, the drop is drastically larger for higher degrees and can reach two orders of magnitude for $l>100$. In the scope of the transport of angular momentum by IGW, $L_p$ should highly influence high degrees and so the SLO below the convective zone (since the wave damping $\tau\propto l^3$), while low degrees that are damped deeper in the star would not be affected.

Another difficulty is the determination of the Brunt-Väisälä frequency at the top of the radiative zone, $N_0$, since the radiative/convective transition is very sharp in stellar evolution codes. In this work, we arbitrarily propose to take the mean value of $N$ over $0.1 H_p$, i.e., $1 \%$ of the thickness of the solar radiative zone, just below the Schwarzschild's criterion. Its value influences more the wave energy spectrum in the case of sharp transition ($\mathcal{F}_r \propto 1/N_0$) than in the case of a smooth transition ($\mathcal{F}_r \propto 1/N_0^{2/3}$), but its influence remains significative. However, uncertainties on $N_0$ does not change the comparison with the turbulence-induced wave energy spectrum since it plays the same role in both models. Indeed, it plays a role only in the transmission of the wave into the radiative zone and not in the excitation process.

The above remark about $N_0$ is also valid for the length $d$ of the thermal transition zone ($r_d \le r \le r_b - L_p$, see \figurename{} \ref{schema situation}). In addition, we have shown the larger $d$, the better the transmission of the waves, so that the wave energy flux at the top of the radiative zone obtained in the case of a discontinuous transition ($d=0$) appears as a lower limit. Several estimates of the thickness of the thermal adjustment layer have theoretically been done in the solar case. \cite{Schmitt1984} found an upper limit at about $500$ km, whereas \cite{Zahn1991} evaluated $d$ to be of the order of the kilometer. More recently, \cite{Rempel2004}, using a semianalytical model of penetrative convection taking into account the interaction with the upflows and a distribution of the plume velocity, estimated $d\approx 350$ km. Given these previous works, we have chosen $d=500$ km in \sectionname{} \ref{result d ne 0} as an upper limit for the transition length. With such a value, the total plume-induced wave energy flux at the top of the radiative zone have been shown to represent about $10 \%$ of the solar flux. Hence, in the case of a linear profile for the Brunt-Väisälä frequency in the transition region, the plume-induced wave energy flux goes from $1 \%$ to $10 \%$ of the solar flux for the considered range $d \in[0,500]$ km. Since $d$ plays the same role in both excitation processes, the turbulence-induced wave energy flux is expected to be between $0.1 \%$ to $1 \%$ of the solar flux for the same range of values for $d$. In the case of a linear smooth transition ($\gamma_d \gg 1$), we have seen that the wave energy flux is proportional to $d^{1/3}$, so that a decrease of one order of magnitude for $d$ induces only a decrease by about a factor two for the wave energy flux. However, the sensitivity of the wave transmission to the parameter $d$ depends on the profile of the transition. For instance, \cite{Lecoanet2013} showed that the wave energy flux is actually proportional to $d$ in the case of a smooth hyperbolic tangent profile for the Brunt-Väisälä frequency at the base of the convective zone. It is thus more strongly sensitive to the value of $d$ than for a linear transition profile. At last, these results could explain the very large wave energy flux observed in numerical simulations. For example, \cite{Dintrans2005} found that IGW could carry up to $40\%$ of the total kinetic energy despite a very low Péclet number at the base of the convective zone (imposed by numerical constraints). As explained in \cite{Dintrans2005}, a low Péclet number means a weak buoyancy braking of the plume (and so an inefficient excitation by penetrative convection), but it also means a very smooth transition \citep[e.g.,][]{Brummell2002} which can considerably enhance the transmission of IGW into the radiative zone. 

Finally, despite the wide range of values considered in \sectionname{} \ref{result d=0} and in this section for the parameters of the excitation model, the same result holds true: excitation by penetrative convection remains more efficient than by turbulent pressure in the Sun and generates a total wave energy flux larger than $0.1\%$ of the solar flux.

\subsection{Limits of the model} 
\label{limits of the model}

It remains to discuss the limits of the model and the improvements to be done in the future. First, in \sectionname{} \ref{sensitivity on input physics}, we have mentioned the dependence of the plume-induced wave energy spectrum on the parameters and the plume characteristics by assuming that they were independent. However, the reality is much more complex. For example, the penetration length has been shown to depend on the filling factor and the plume velocity at the base of the convective zone. \cite{Schmitt1984} and, later, \cite{Zahn1991} demonstrated that $L_p$ is in fact proportional to $\mathcal{A}^{1/2} V_b^{3/2}$. Besides, we can intuitively understand that the plume radius and the plume lifetime must be correlated. The larger the plume, the longer its lifetime. For instance, the characteristic destruction time of the plume by baroclinic eddies is proportional to the plume radius from the model of \cite{Jones1997} in \eqref{t_res}.

Second, the plume characteristics are not unique and must be distributed around mean values. This is contradictory with the assumption that all the plumes are identical. For example, a distribution of plume velocities at the base of the convective zone could affect the penetration length and the steepness of the thermal transition \citep{Rempel2004}. A distribution of the plume parameters could also have an impact on the shape of the wave energy spectrum at the top of the radiative zone. Numerical simulations of convection will be essential to answer these questions. Exhaustive studies could be done by varying parameters such as the Reynold's or the Péclet number of the simulations in order to find scaling relations and to extrapolate to stellar regimes.

Lastly, we have found that the total wave energy flux at the top of the radiative zone represents one fourth of the mean plume kinetic energy flux in the case of a smooth thermal transition ($d=500$ km). However, this result has to be taken with caution since it accounts for very low frequencies at which the present description is not valid. For example, physical processes like non-adiabatic effects cannot be neglected for frequencies that are typically lower than $10^{-6} \mbox{ rad~s}^{-1}$. In addition, the derived wave transmission coefficient diverges when $\omega \rightarrow 0$ in the case of a very smooth transition (see \figurename{} \ref{ratio smooth sharp}). As a consequence, the amplitude of the transmitted low-frequency waves rises steeply and the linear approximation is no more valid as soon as $k_r \xi_r \approx 1$ and non-linear processes like wave breaking can occur. Finally, while beyond the scope of this article, we note that the Coriolis acceleration is to be taken into account in the modeling of the driving processes for wave frequencies around (or lower) than twice the internal rotation rate. This is however a challenging task since it requires a full two-dimensional description of both the wave field and the plume dynamics.
%

%
\section{Conclusion and perspectives}
%
\label{conclusion}

In this paper, we propose a semianalytical model of excitation of IGW by penetrative plumes. We consider the pressure exerted  by an ensemble of plumes at the radiative/convective interface as the driving mechanism that is supposed to be random and stationary.
The wave equation with source term is solved by taking into account the thermal transition from a convective to a radiative gradient at the base of the convective zone. We then determine the amplitude of the waves and investigate the effect of the steepness of the thermal adjustment layer on the wave transmission into the radiative zone. The description of the plumes at the base of the upper convective region follows the model of \cite{Rieutord1995}. We thus obtain the expression of the wave energy flux per unit of frequency in the whole radiative zone in \eqref{spectral wave energy flux} as well as its approximated version in \eqref{luminosity approximate}. Finally, the model depends on four free parameters: the filling factor, the plume lifetime, the length of penetration and the one of the thermal transition layer.

Numerical calculations for a solar model with reasonable values for the parameters show that the excitation by penetrative convection is five to ten times more efficient than the driving by turbulence in the convective zone as formulated by \cite{Kumar1999}. Moreover, we demonstrate that plume-induced waves are able to modify the internal rotation profile of the Sun in about 0.1~Gyr, which is more than ten times faster than with turbulence-induced waves. Plume-induced waves are thus able to considerably affect the internal rotation rate of solar-like stars on times shorter than their lifetime on the main-sequence. Furthermore, we show that these results are conservative despite a wide range of values considered for the parameters of the model.

These preliminary results on the Sun will have to be numerically quantified including transport by meridional circulation and shear-induced turbulence with plume-induced waves and turbulence-induced waves. The study will have to be extended to subgiant and red giant stars with appropriate values of the free parameters. The present excitation model enables an easy implementation in 1D stellar evolution codes, and could be used to investigate the redistribution of angular momentum all along the evolution of stars taking into account transport by plume-induced internal gravity waves as well as turbulence-induced ones, and other mechanisms.

At last, the effect of rotation on the wave dynamics will have to be included in order to get a more realistic view of the transport by internal waves in stars. Indeed, whatever the driving process is, low frequency waves are affected by the Coriolis force. Thus, it cannot be neglected and it will have an impact on the transport by gravito-inertial waves in the radiative zone \citep{Mathis2004,Mathis2009} and on the excitation process \citep[e.g.,][]{Mathis2014}. Such a development represents a theoretical challenge for the future since we will have to go beyond the traditional approximation for sub-inertial waves (i.e., $\omega \le 2 \Omega$) and to tackle the problem in 2D, as a function of the radial and latitudinal variables. 
%



%
\bibliographystyle{aa}
\bibliography{bib}
%

\appendix

%
\section{Plume lifetime and restratification in the Sun}
%

\label{restratification}

\begin{figure}[t]
\centering
\includegraphics[scale=0.18, trim= 0cm 0cm 0cm 0cm, clip]{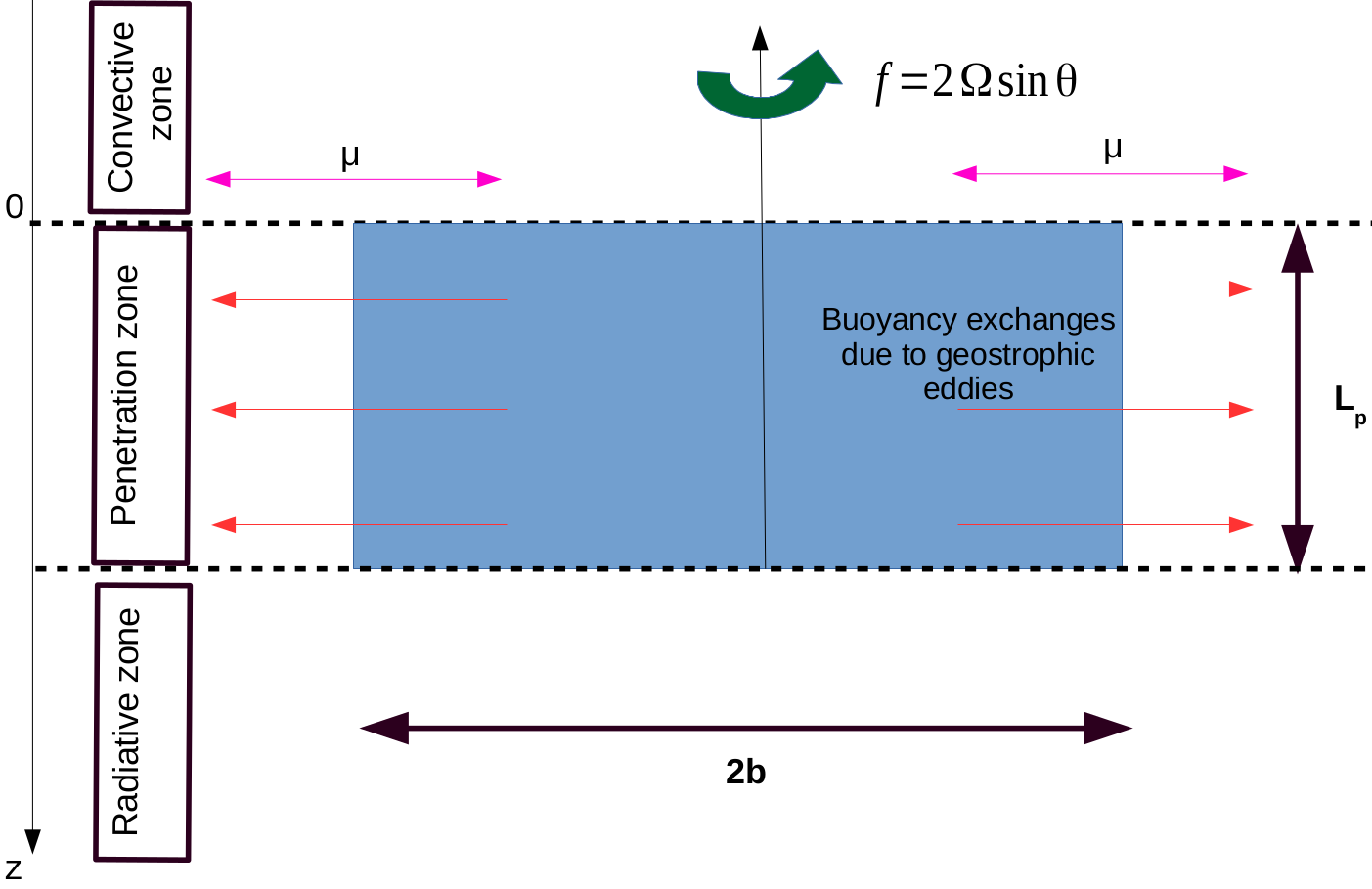} 
\caption{Schematic view of the plume (in blue) in the penetration region, surrounded by a stable stratitfied medium. $\mu$ represents the horizontal length scale of the density contrast between the two distinct parts.}
\label{schema_restratification}
\end{figure}

We adapt the model developed by \cite{Jones1997} in the penetration zone of the Sun where a lateral density contrast exists between the plume and the surroundings. We consider a fluid cylinder of height and radius equal to $L_p$ and $b$, respectively (see \figurename \ref{schema_restratification})

\subsection{Geostrophic wind in the solar penetration zone}

\cite{Jones1997} investigated the restratification of convective patches in the oceans for which the geostrophic approximation is valid. Under these conditions, the flows follow the geostrophic wind equations resulting in a balance between the pressure gradient and the Coriolis force. In order to adopt their model to the Sun, we must first verify the validity of the geostrophic approximation in the solar penetration zone.
\begin{enumerate}

\item Acceleration in the corotating frame with the base of the convection zone must be negligible compared to the Coriolis force, so that the Rossby number 
\begin{equation}
R_o=\frac{U}{2 \Omega L}
\label{rossby_number}
\end{equation}
is smaller than unity, with $\Omega$ the mean angular rotation rate, $L$ the characteristic horizontal length scale and $U$ the zonal velocity in the corotating frame. In the solar tachocline case, taking $\Omega \approx 2.5$ $10^{-6} \mbox{ rad~s}^{-1}$, $L \approx 2 b\approx 2$ $10^4$ km, we will verify a posteriori that $R_o \ll 1$ provided that $U   \ll 100$~m~s${}^{-1}$.

\item Viscous stress can be neglected compared to the Coriolis term in the equation of motion, the Ekman number
\begin{equation}
E_k=\frac{\nu}{2 \Omega L^2} \mbox{  ,}
\label{Ekman_number}
\end{equation}
with $\nu$ the viscosity, is much smaller than unity. Even with a turbulent viscosity due to local shear expected to be around $\nu_t \sim 10^6 \mathrm{cm}{}^{2}.\mbox{s}^{-1}$, the Ekman number stays very small.

\item Finally, the vertical length scale must be negligible compared to horizontal one. In the solar case, taking $L_p\approx 0.1 H_p$, $H_p \approx 5.10^4$ km, the pressure scale height at the base of the convective zone, and $b\approx 10^4$ km, the ratio is equal to 
\begin{equation}
\frac{L_p}{2b} \approx 10^{-1} \mbox{  .}
\end{equation} 

\end{enumerate}
Thus, the geostrophic approximation in the solar penetration zone will be valid on condition that the zonal wind in the corotating frame is slow enough; this will be verified a posteriori.

\subsection{Plume destruction time scale by geostrophic eddies \citep{Jones1997}}

By assuming vertical hydrostatic balance (which is possible inside the plume if we consider an almost uniform and stationary flow), we can then use the thermal wind equation
\begin{align}
\vec{u}&=\frac{1}{f \rho} \vec{k} \times \boldsymbol{\nabla}_h p \nonumber \\
\frac{\partial \vec{u}}{\partial z}&=\frac{g}{f \rho} \vec{k} \times \boldsymbol{\nabla}_h \rho \mbox{  ,}
\end{align}
where $f=2 \sin\theta \Omega$ is the Coriolis parameter, $\rho$ the density, $p$ the pressure and $\vec{k}$ the unit vector perpendicular to the surface of the sphere. 
The density contrast $\delta \rho$ between the plume and the surrounding medium creates a vertical shear, which obeys to
\begin{equation}
\frac{\partial u}{\partial z} \approx \frac{g }{f \mu}\frac{\delta \rho}{\rho}=-\frac{1}{f\mu} \delta b \mbox{  ,}
\label{vent_zonal}
\end{equation}
where $\mu$ is the horizontal length scale on which density exchanges take place and $\delta b=-g \delta \rho/\rho$ represents buoyancy acceleration.
To go further, we use a first order expansion at $z\sim L_p$ of $\delta b$,
\begin{equation}
\delta b \approx (\delta b)_{L_p}+\left( \frac{d\delta b}{dz}\right)_{L_p}(z-L_p)=-N_t^2 L_p (\frac{z}{L_p}-1) \mbox{  ,}
\label{buoyancy_ordre1}
\end{equation}
where we impose $(\delta b)_{L_p}=0$ and we note 
\begin{equation}
\left( \frac{d\delta b}{dz}\right)_{L_p}=-N_t^2 \mbox{  ,}
\end{equation}
with $N_t^2$ the mean Brunt-Väisälä frequency in the penetration zone. By injecting \eqref{buoyancy_ordre1} in \eqref{vent_zonal}, we obtain the zonal wind as a function of the altitude
\begin{equation}
u=\frac{N_t^2 L_p^2}{f \mu}\left( \frac{z^2}{2 L_p^2}-\frac{z}{L_p}+\frac{1}{3}\right) \mbox{  ,}
\end{equation}
where the constant of integration is chosen to satisfy $\int_0^{L_p} u(z)\mathrm{d}z=0$. 
\cite{Jones1997} showed that the thermal wind undergoes a baroclinic instability when the radius of the fluid column is longer than the Rossby length scale of deformation
\begin{equation}
L_D \approx \frac{N_t L_p}{f} \mbox{  .}
\label{Rossby_radius}
\end{equation}
In the penetration zone, injection of convective material imposes a quasi-adiabatic gradient before reaching a radiative regime where the Brunt-Väisälä frequency reaches $N_0 \approx 300$ $\mu H_z$. Hence, we can estimate $N_t\sim1-10$ $\mu H_z$. With $L_p \sim 0.1 H_p$, we find $L_D \approx 10^2-10^3$ km $ \le b$ and the fluid column in the Sun undergoes a baroclinic instability.
Moreover, at $z=0$ , we find
\begin{equation}
u(z=0)=\frac{N_tL_p}{3}\frac{L_D}{\mu} \mbox{  .}
\end{equation}
We then assume that the density exchanges take place on a more expanded length scale than the Rossby radius of deformation, so $\mu \ge  L_D$ . Therefore, $u(z=0) \le 10$~m~s$^{-1}$, the Rossby number $R_o \ll 1$ and we verify  a posteriori the thermal wind approximation (cf. paragraph on Rossby number \eqref{rossby_number}).

By conservation of potential buoyancy, $\overline{\delta b}$ is conserved in the cylinder of fluid, where the bar denotes time average over a characteristic time scale. Let $\vec{v}$ be the eddies-induced velocity field. The Reynolds transport theorem enables us to write
\begin{equation}
\frac{\partial }{\partial t} \left(\int \limits_{cylinder} \overline{(\delta b)}\hspace{0.1cm} \mathrm{d}V\right)=2 \pi b \int_{0}^{L_p} \overline{(\delta b)\hspace{0.1cm}\vec{v} }\cdot \vec{n} \mathrm{d}z \mbox{  ,}
\label{conservation_buoyancy}
\end{equation}
where $\vec{n}$ is the unit vector perpendicular to the lateral surface and where we have neglected the contribution of the buoyancy fluxes through the top and the bottom of the cylinder. To go further, we suppose
\begin{equation}
\overline{(\delta b)\hspace{0.1cm} \vec{v}} \cdot \vec{n}=c_e (\delta b) u_{z=0} \mbox{  ,}
\label{efficacite_eddy}
\end{equation}
with $c_e$ representing the efficiency coefficient of the transport by geostrophic eddies. The right-hand side of equation \eqref{efficacite_eddy} becomes after integration
\begin{equation}
2 \pi b \int_{0}^{L_p} \overline{(\delta b)\hspace{0.1cm}\vec{v} }\cdot \vec{n} \mathrm{d}z=2 \pi b\frac{c_e N_t^3L_p^3}{6} \frac{L_D}{\mu} \mbox{  .}
\end{equation}
The left-hand side of equation \eqref{efficacite_eddy} gives
\begin{equation}
\frac{\partial }{\partial t} \left(\int \limits_{cylinder} \overline{(\delta b)}\hspace{0.1cm} \mathrm{d}V\right)\approx \frac{1}{\tau_{res}} \frac{\pi b^2 N_t^2 L_p^2}{2} \mbox{  ,}
\end{equation}
whence we can deduce the restratification time scale
\begin{equation}
\tau_{res} \approx \frac{3}{2}\frac{b}{c_e N_t L_p} \frac{\mu}{L_D} \mbox{  .}
\label{t_res}
\end{equation}
From numerical experiments and observations, \cite{Jones1997} derived $c_e=0.027$ in the case of the oceans. Given the lack of knowledge about this process in the Sun, we use this value to estimate the plume lifetime in our case. We find $t_{res} \sim 10^6-10^7$ s with $\mu \sim L_D$, i.e $\nu_p=1/\tau_p \sim 0.1 -1 \hspace{0.1cm} \mu H_z$. Note that this value is of the same order of magnitude than the turnover convective frequency, $\omega_c=v_c/l_c$, as given by the MLT in a standard model, with $\mathrm{v}_c$ the convective velocity and $l_c$ the mixing length.

%
\section{Derivation of the wave energy flux}
%

\label{derivation of the wave energy flux}

In the following, we detail the derivation of the wave power spectrum in the radiative zone. We will focus on the pressure due to convective plumes as driving mechanism in the penetration zone, and we will take into account the effect of the thermal transition on the transmission of the generated waves, as illustrated in \figurename{} \ref{schema situation}.

\subsection{Adiabatic wave equation with source term}

\label{adiabatic wave equation}

As explained in \sectionname{} \ref{wave equation and source term}, the total velocity field is split into a wave component $\vec{v}$ and a plume component $\vec{V_p}$. We neglect the non-linear wave term $\boldsymbol{\nabla}\cdot(\rho \vec{v} \otimes \vec{v} )$ and the plume-related coupling terms, like $\boldsymbol{\nabla}\cdot(\rho \vec{V_p} \otimes \vec{v} )$. We also assume that the time variation of the plume impulsion in the penetration zone, $\partial_t (\rho \vec{V}_p) $,
does not participate into the excitation of the wave, and that the plume destruction is mainly due to instabilities and turbulent motions in this region. The equations of dynamics are then given by \eqref{momentum eq} and \eqref{continuity eq}.

To go further, we adapt the procedure used by \cite{Unno1989} in order to include the forcing term and the continuous frequency spectrum of the plume-induced wave packet. 
We introduce the Lagrangian wave displacement vector $\boldsymbol{\xi}(\vec{r},t)$ related to the wave velocity field by $\vec{v}=\partial_t \boldsymbol{\xi}$.
The Eulerian perturbations are decomposed onto the spherical harmonics, like the wave velocity field in \eqref{decomposition wave velocity} or, for example, the radial Lagrangian displacement
\begin{equation}
\xi_r(\vec{r},t)=\sum_{l,m} \xi_{r,l,m}(r,t) Y_l^m(\theta,\phi )\mbox{  .}
\label{xi_r serie}
\end{equation}
Deriving the spectral density of the wave specific energy \eqref{spectral density ensemble} requires the computation of the time Fourier transform of the wave velocity field. By taking the time Fourier transform of \eqref{momentum eq} and \eqref{continuity eq} and using the fact that spatial differential operators commute with time Fourier transform,
we obtain
\begin{align}
-\omega^2 \boldsymbol{\hat{\xi}}+\frac{\boldsymbol{\nabla} \hat{p^\prime}}{\rho} - \frac{\hat{\rho^\prime}}{\rho} \vec{g} &=
-\mathcal{TF}\left[\frac{1}{\rho} \boldsymbol{\nabla} \cdot (\rho \vec{V}_p\otimes \vec{V}_p)\right]\label{momentum tf}\\
 \frac{\delta \hat{\rho}}{\rho}+\boldsymbol{\nabla} \cdot \boldsymbol{\hat{\xi}} &=0 \mbox{  .} \label{continuity tf}
\end{align}
The horizontal part of \eqref{momentum tf} enables to link the horizontal displacement vector $\hat{\boldsymbol{\xi}}_{h,l,m}$ to $\hat{p}^{\prime}_{l,m}$ and the horizontal part of the forcing term
\begin{equation}
\boldsymbol{\hat{\xi}_{h}}=\frac{1}{\omega^2} \left\{\boldsymbol{\nabla}_{\perp}\left( \frac{\hat{p}^{\prime}}{\rho}\right)+
\mathcal{TF}\left[\frac{1}{\rho} \left(\boldsymbol{\nabla} \cdot (\rho \vec{V}_p\otimes \vec{V}_p)\right)_{\perp}\right]\right\}
\mbox{  ,} \label{horizontal displacement tf}
\end{equation}
with $\boldsymbol{\nabla}_{\perp}$ the horizontal nabla operator.
To continue, we proceed in the same way than \citet[Eq. 13.40]{Unno1989}. We compute the horizontal divergence of \eqref{momentum tf} and replace $\boldsymbol{\nabla}_{\perp} \cdot\hat{\boldsymbol{\xi_h}}$ from \eqref{continuity tf} to find
\begin{equation}
\frac{\delta \hat{\rho}}{\rho}+\frac{1}{r^2}\frac{\partial}{\partial r}\left( r^2 \hat{\xi}_r \right)+
\boldsymbol{\nabla}_{\perp}^2 \left( \frac{ \hat{p^\prime}}{\rho\omega^2}\right)
=- \frac{1}{\omega^2} \boldsymbol{\nabla}_{\perp} \cdot \left\{ \mathcal{TF}\left[\frac{1}{\rho} \boldsymbol{\nabla} \cdot (\rho \vec{V}_p\otimes \vec{V}_p)\right] \right\} \label{div hor continuity}
\end{equation}
Using the adiabatic closure hypothesis (in Fourier space),
\begin{equation}
\delta \hat{\rho} =\delta\hat{ p} /c^2=(\hat{p}^\prime-\rho g\hat{\xi}_r)/c^2 \mbox{  ,}
\label{adiabatic closure}
\end{equation}
and the decomposition onto spherical harmonics \eqref{xi_r serie}, \eqref{div hor continuity} can be rewritten 
\begin{align}
\sum_{l,m}&\left[\frac{1}{r^2}\frac{d}{dr} \left( r^2 \hat{\xi}_{r,l,m} \right) - \frac{g}{c^2} \hat{\xi}_{r,l,m} + \left(1-\frac{S_l^2}{\omega^2}\right)  \frac{\hat{p}^\prime_{l,m}}{\rho c^2} \right] Y_l^m(\theta,\phi ) \nonumber\\
&=-\frac{1}{\omega^2} \boldsymbol{\nabla}_{\perp} \cdot \left\{ \mathcal{TF}\left[\frac{1}{\rho} \boldsymbol{\nabla} \cdot (\rho \vec{V}_p\otimes \vec{V}_p)\right] \right\}\mbox{  .}
\label{equation continuity momentum}
\end{align}
Similarly, using $\hat{\rho}^\prime=\delta \hat{\rho} - \xi_r \boldsymbol{\nabla} \rho$ and \eqref{adiabatic closure}, the radial part of \eqref{momentum tf} becomes
\begin{align}
\sum_{l,m}&\left[ \frac{1}{\rho} \frac{ d \hat{p}^\prime_{l,m}}{d r} + \frac{g}{\rho c^2}  \hat{p}^\prime_{l,m} + (N^2-\omega^2) \hat{\xi}_{r,l,m}\right]Y_l^m(\theta,\phi ) \nonumber\\
&=-\vec{e}_r \cdot \mathcal{TF}\left[\frac{1}{\rho} \boldsymbol{\nabla} \cdot (\rho \vec{V}_p\otimes \vec{V}_p)\right] \mbox{  ,}
\end{align}
with $c$ the sound speed, $S_l$ the Lamb frequency and $N$ the Brunt-Väisälä frequency.
After  projecting onto $Y_l^m(\theta,\phi )$, we obtain a first-order linear system of two differential equations for $\hat{\xi}_{r,l,m}(r,\omega)$ and $\hat{p}^\prime_{l,m}(r,\omega)$ 
\begin{align}
\frac{1}{r^2}\frac{d}{dr} \left( r^2 \hat{\xi}_{r,l,m} \right) - \frac{g}{c^2} \hat{\xi}_{r,l,m} + \left(1-\frac{S_l^2}{\omega^2}\right)  \frac{\hat{p}^\prime_{l,m}}{\rho c^2}  &=F_1\label{eq F1}\\
\frac{1}{\rho} \frac{ d \hat{p}^\prime_{l,m}}{d r} + \frac{g}{\rho c^2}  \hat{p}^\prime_{l,m} + (N^2-\omega^2) \hat{\xi}_{r,l,m}
&=F_2 \label{eq F2} \mbox{  .}
\end{align}
The differential system is similar to Eq. (14.2) and (14.3)  in \cite{Unno1989}, with the additional right-hand side source terms $F_1$ and $F_2$, given by
\begin{align}
F_1&= -\int \frac{1}{\omega^2} \boldsymbol{\nabla}_{\perp} \cdot \left\{ \mathcal{TF}\left[\frac{1}{\rho} \boldsymbol{\nabla} \cdot (\rho \vec{V}_p\otimes \vec{V}_p)\right] \right\}\overline{Y_l^m}\mathrm{d} \Omega \label{F1}\\
F_2&= -\int \vec{e}_r \cdot \mathcal{TF}\left[\frac{1}{\rho} \boldsymbol{\nabla} \cdot (\rho \vec{V}_p\otimes \vec{V}_p)\right] \overline{Y_l^m}\mathrm{d} \Omega \label{F2} \mbox{  ,}
\end{align}
with the solid angle given by $\mathrm{d}\Omega=\sin\theta \mathrm{d} \theta \mathrm{d} \phi$. 

\subsection{ Modeling of the source term in the wave equation}
\label{modelling stress}
Considering the velocity profile in \eqref{profil_Vp}, the pressure gradient due to a single plume is given by
\begin{equation}
\boldsymbol{\nabla} \cdot (\rho \vec{V}_p\otimes \vec{V}_p)=e^{-2\nu_p^2 t^2 }\left[ \frac{\partial}{\partial r} +\frac{2}{r} \right] \left(\rho V_0^2
e^{-S_h^2/b^2}\right)\hspace{0.1cm} \vec{e}_r \mbox{  .} \label{div pressure}
\end{equation}
We assume that the radius at the interface, $r_b$, is much larger than the penetration length, $L_p \ll r_b$. Hence, the operator $2/r$ can be neglected compared to $\partial_r$, and the radius is taken constant in this region. Therefore, using \eqref{div pressure}, \eqref{F2} can be written as
\begin{equation}
F_2=-\frac{\alpha(\omega)}{\rho} \beta_l^m\frac{d}{d r} \left( \rho V_0^2 \right)  \mbox{  ,}
\label{F_2 expression}
\end{equation}
where we have defined
\begin{align}
\alpha(\omega)&= \mathcal{TF}\left[e^{-2\nu_p^2 t^2 }\right]=\sqrt{\frac{\pi}{2}}\frac{e^{-\omega^2/8 \nu_p^2 }}{\nu_p} \label{alpha omega} \\
\beta_l^m&=\int e^{-s_b^2/b^2} \overline{Y_l^m}\mathrm{d} \Omega=\beta_l^m(\theta_0,\phi_0) \label{beta l_m annexe}\mbox{  ,}
\end{align}
with $s_b=S_h(r_b,\theta,\phi;\theta_0,\phi_0)$ following \eqref{def S_h}. We also find that \eqref{F1} cancels, $F_1=0$, since \eqref{div pressure} is collinear with the radial direction.
For the same reason, using \eqref{div pressure} in \eqref{horizontal displacement tf}, we see that the toroidal part of the displacement vector is null, $\xi_{T,l,m}=0$, and that the poloidal part, renamed $\hat{\xi}_{h,l,m}$, is linked to the perturbation of pressure through the expression
\begin{equation}
\hat{\xi}_{P,l,m}(r,\omega)=\hat{\xi}_{h,l,m}(r,\omega)=\frac{\hat{p}_{l,m}^\prime(r,\omega)}{\rho r \omega^2} \mbox{  .}
\label{xi_h vs p}
\end{equation}
Finally, \eqref{eq F1} and \eqref{eq F2} can be rewritten in the form of the following non-homogeneous first-order differential system 
\begin{equation}
\frac{d\vec{z}}{d r}(r,\omega)=\vec{A}(r,\omega)\vec{y}(r,\omega)+\vec{b}(r,\omega) \mbox{  ,}
\label{differential system y}
\end{equation}
where $\vec{z}$, \vec{b} and $\vec{A}$ are given by \eqref{z and b} and \eqref{matrix A}.

\setcounter{subsubsection}{0}
\subsection{Homogeneous equation}

To solve the homogeneous differential system, i.e., \eqref{differential system y} with $\vec{b}=0$, we use the change of variables \citep{Unno1989}
\begin{align}
\mathrm{v}_{l,m}(r)&=\rho^{1/2} r c_s\left|\frac{S_l^2}{\omega^2}-1\right|^{-1/2} \hat{\xi}_{r,l,m}\label{v_r}\\
\mathrm{w}_{l,m}(r)&=\rho^{-1/2}r \left|N^2-\omega^2\right|^{-1/2} \hat{p}_{l,m}^\prime \label{w_h} \mbox{  .}
\end{align}
This leads to a more tractable second-order differential linear equation for $\mathrm{v}_{l,m}$
\begin{align}
\frac{d^2 \mathrm{v}_{l,m}}{dr^2}+k_{r}^2 \mathrm{v}_{l,m} = 0 \mbox{  ,}
\label{eq v_r}
\end{align}
with $k_r$ given by \eqref{k_r} and where we have neglected the term $f(P)$ in Eq. (16.11) of \cite{Unno1989}. Therefore, $\hat{\xi}_{r,l,m}$ is obtained through \eqref{v_r} by solving \eqref{eq v_r}, and $\hat{\xi}_{h,l,m}$ is deduced from \eqref{xi_h vs p}, \eqref{w_h} and the relation
\begin{align}
\mathrm{w}_{l,m}(r)\approx\mathrm{sgn}(S_l^2-\omega^2) |k_{r}|^{-1} \frac{d \mathrm{v}_{l,m}}{dr} \label{w_par_v}
\end{align}
which comes from \eqref{equation continuity momentum}. This approach will be used in the following paragraphs to find the two linearly independent solutions (subscript $1$ or $2$) of the homogeneous differential system for each region: the quasi-adiabatic region, composed of the penetration zone and the convective zone, the transition region and the radiative interior (see \figurename \ref{schema situation}).
 
\addtocounter{subsubsection}{1} 
\subsubsection*{\Alph{section}.\arabic{subsection}.\arabic{subsubsection} Quasi-adiabatic region $(r_b-L_p \le r)$}

In the convective zone and the penetration zone, we assume that the temperature gradient is quasi-adiabatic, $N^2\sim 0$ and the waves are evanescent. In this region, the WKB solutions of \eqref{eq v_r} are a good approximation. The wave functions, $\vec{z}_{1}^a$ and $\vec{z}_{2}^a$, in the quasi-adiabatic region (superscript $a$) are then estimated up to a constant by
\begin{align}
\left( \:\begin{array}{c}
z_{1,r}^a\vspace{0.2cm}\\
z_{1,h}^a
\end{array} \:\right)
&=i \rho^{-1/2} r^{-3/2}\left(\frac{r}{r_b-L_p}\right)^{\Lambda}
\left( \:\begin{array}{c}
\Lambda^{1/2}\vspace{0.2cm}\\
\Lambda^{-1/2}
\end{array} \:\right)\label{y_1^a}\\
\left( \:\begin{array}{c}
z_{2,r}^a\vspace{0.2cm}\\
z_{2,h}^a
\end{array} \:\right)
&=i\rho^{-1/2} r^{-3/2}\left(\frac{r_b-L_p}{r}\right)^{\Lambda}
\left( \:\begin{array}{c}
\Lambda^{1/2}\vspace{0.2cm}\\
-\Lambda^{-1/2}
\end{array}\:\right) \label{y_2^a} \mbox{  ,}
\end{align}
where the subscripts $r$ and $h$ refer to the radial and horizontal components of the vectors, $r_b$ is the radius of the base of the convective zone (as prescribed by the Schwarzschild's criterion), $L_p$ is the penetration length and $\Lambda=\sqrt{l(l+1)}$.

\addtocounter{subsubsection}{1}
\subsubsection*{\Alph{section}.\arabic{subsection}.\arabic{subsubsection} Radiative interior $(r \le r_d)$}

In the same way, in the radiative zone (superscript $r$), propagative waves are estimated with the WKB solutions of \eqref{eq v_r}. The inward and outward wave functions, $\vec{z}_{1}^r$ and $\vec{z}_{2}^r$, are respectively given up to a constant by

\begin{align}
\left( \:\begin{array}{c}
z_{1,r}^r\vspace{0.2cm}\\
z_{1,h}^r
\end{array} \:\right)
&=i\rho^{-1/2} r^{-3/2}\exp\left(- i \int_r^{r_d} k_r \mathrm{d}r\right)
\left( \:\begin{array}{c}
\Lambda^{1/2}\left( \frac{N^2}{\omega^2}-1\right)^{-1/4}\vspace{0.2cm}\\
i\Lambda^{-1/2} \left( \frac{N^2}{\omega^2}-1\right)^{1/4}
\end{array} \:\right)\label{y_1^r}\\
\left( \:\begin{array}{c}
z_{2,r}^r\vspace{0.2cm}\\
z_{2,h}^r
\end{array} \:\right)
&=i\rho^{-1/2} r^{-3/2}\exp\left(+ i \int_r^{r_d} k_r \mathrm{d}r\right)
\left( \:\begin{array}{c}
\Lambda^{1/2}\left( \frac{N^2}{\omega^2}-1\right)^{-1/4}\vspace{0.2cm}\\
-i\Lambda^{-1/2} \left( \frac{N^2}{\omega^2}-1\right)^{1/4}
\end{array}\:\right)
\end{align}
where $r_d$ represents the radius at the top of the radiative zone $r_d=r_b-L_p-d$.

\addtocounter{subsubsection}{1}
\subsubsection*{\Alph{section}.\arabic{subsection}.\arabic{subsubsection} Thermal transition layer $(r_d \le r \le r_b - L_p)$}

Once the plume is slowed down enough, radiative diffusion can operates efficiently. In this so-called thermal adjustment layer, the temperature gradient undergoes a transition from a quasi-adiabatic gradient to a radiative one. We follow here the work of \cite{Lecoanet2013} and assume that the Brunt-Väisälä frequency varies linearly in the layer (\figurename \ref{schema situation})
\begin{equation}
N^2=-\frac{N_0^2}{d}\left(z^\prime - \frac{d}{2} \right) \mbox{  ,}
\label{def N^2}
\end{equation}
with $z^\prime=r-(r_b-L_p)+d/2$. By injecting \eqref{def N^2} in the expression of the radial wavenumber \eqref{k_r}, \eqref{eq v_r} becomes:
\begin{equation}
\frac{d^2 \mathrm{v}_{l,m}}{dz'^2}+ \left( -\frac{k_h^2 N_0^2}{ \omega^2 d} z^\prime + \frac{k_h^2 N_0^2}{\omega^2 d} \frac{d(N_0^2-2 \omega^2)}{2 N_0^2} \right) \mathrm{v}_{l,m} = 0 \mbox{  .}\label{eq v_r_developpee}
\end{equation}
As in \cite{Lecoanet2013}, we define
\begin{align}
K_1&=\frac{k_h^2 N_0^2 }{ d \omega^2}\\
z_t&=\frac{d (N_0^2-2 \omega^2)}{2 N_0^2} \mbox{  ,}
\end{align}
with $z_t$ the wave turning point (where $N^2=\omega^2$) and $k_h \approx \Lambda / r_b$ the horizontal wavelength, supposed to be constant in the transition layer.
Then, using the change of variable $\chi=K^{1/3}(z^\prime-z_t)$, we find that \eqref{eq v_r_developpee} verifies the equation of Airy:
\begin{equation}\frac{d^2 \mathrm{v}_{l,m}}{d \chi^2}-\chi \mathrm{v}_{l,m}=0 \mbox{  ,} \end{equation}
whose the solution is a linear combination of the Airy functions $A_i(\chi)$ and $B_i(\chi)$.
To determine the horizontal part $\hat{\xi}_{h,l,m}$, we use \eqref{w_par_v} that can be written as a function of the variable $\chi$
\begin{equation}
\mathrm{w}_{l,m}=\frac{1}{K_1^{1/3}|\chi|^{1/2}} \frac{d \chi}{dr}\frac{d \mathrm{v}_{l,m}}{d \chi}=
\frac{1}{|\chi|^{1/2}}\frac{d \mathrm{v}_{l,m}}{d \chi} \mbox{  .}
\end{equation}
Therefore, the two linearly independent\footnote{The Wronskian $W$ of $\vec{z}_1^t$ and $\vec{z}_2^t$ is equal to 
\[W=-\rho^{-1} r^{-3}K_1^{1/3} \left( A_i(\chi) \frac{d B_i}{d \chi} (\chi)-B_i(\chi)\frac{d A_i}{d \chi}(\chi)\right)=-\rho^{-1}r^{-3} K_1^{1/3}\frac{1}{\pi}\] which is a non-zero value whatever the radius is, meaning that they are well linearly independent.} solutions of the homogeneous wave equation, $\vec{z}_{1}^t$ and $\vec{z}_{2}^t$, in the transition layer (superscript $t$) read up to a constant
\begin{align}
\left( \:\begin{array}{c}
z_{1,r}^t\vspace{0.2cm}\\
z_{1,h}^t
\end{array} \:\right)
&=i\rho^{-1/2} r^{-2}
\left( \:\begin{array}{c}
\sqrt{l(l+1)}A_i(\chi)\vspace{0.2cm}\\
\frac{K_1^{1/3}}{k_h} \frac{d A_i}{d \chi} (\chi)
\end{array} \:\right)\\
\left( \:\begin{array}{c}
z_{2,r}^t\vspace{0.2cm}\\
z_{2,h}^t
\end{array} \:\right)
&=i\rho^{-1/2} r^{-2}
\left( \:\begin{array}{c}
\sqrt{l(l+1)}B_i(\chi)\vspace{0.2cm}\\
\frac{K_1^{1/3}}{k_h} \frac{d B_i}{d \chi} (\chi)
\end{array}\:\right) \mbox{  .}
\end{align}

\setcounter{subsubsection}{0}
\subsection{Particular solution of the non-homogeneous system}

We suppose that wave driving occurs exclusively in the penetration zone where buoyancy braking is the strongest (i.e., $\vec{b} \ne 0$ for $r_b - L_p<r<r_b$). We derive a particular solution of the non-homogeneous system thanks to the method of variation of parameters. In the quasi-adiabatic region $(r_b \le r)$, we search a solution in the form of
\begin{equation}
\vec{z}_0^a(r)=\mu(r) \vec{z}_1^a(r) + \lambda(r)\vec{z}_2^a(r)  \mbox{  .}
\label{y_0}
\end{equation}
By injecting \eqref{y_0} in \eqref{differential system y}, the functions $\mu$ and $\lambda$ must verify the linear system 
\begin{equation}
\frac{d \mu}{dr} \vec{z}_1^a+\frac{d  \lambda}{dr} \vec{z}_2^a=\vec{b}
\end{equation}
 whose the solution is obtained by using the Cramer's rule 
\begin{align}
\frac{ d \mu}{dr}&=\frac{- z_{2,r}^a b}{z_{1,r}^az_{2,h}^a-z_{2,r}^az_{1,h}^a}\\
\frac{d\lambda}{dr}&=\frac{ z_{1,r}^a b}{z_{1,r}^az_{2,h}^a-z_{2,r}^az_{1,h}^a} \mbox{  ,}
\end{align}
where $b=i F_2/r \omega$. From \eqref{F_2 expression}, \eqref{y_1^a} and \eqref{y_2^a}, integration leads to 
\begin{align}
\mu(r)&=\frac{\alpha(\omega)}{2\omega} \beta_{l,m} \Lambda^{1/2}\int_r^{r_b}\frac{d}{dr}\left( \rho V_0^2\right)
\rho^{-1/2} r^{1/2} \left( \frac{r_b-L_p}{r}\right)^{\Lambda}  \mathrm{d} r \label{coeff mu}\\
\lambda(r)&=-\frac{\alpha(\omega)}{2\omega} \beta_{l,m} \Lambda^{1/2}\int_r^{r_b} \frac{d}{dr}\left( \rho V_0^2\right)
\rho^{-1/2} r^{1/2} \left( \frac{r}{r_b-L_p}\right)^{\Lambda}\mathrm{d} r\label{coeff lambda}
\end{align}
where $\alpha(\omega)$ and $\beta_l^m$ are given by \eqref{alpha omega} and \eqref{beta l_m annexe}, and where we have arbitrarily chosen $\mu(r_b)=\lambda(r_b)=0$.

\subsection{General solution and boundary conditions}

The general wave function is obtained by ensuring the continuity of the radial and the horizontal displacements at the limits between each region. We have also to consider two boundary conditions which are taken at surface and at the center of the star. In the quasi-adiabatic region $(r_b-L_p\le r)$, the general solution $\vec{z}$ reads
\begin{equation}
\vec{z}=A_1\vec{z}_1^a+A_2\vec{z}_2^a+\vec{z}_0^a(r) \mbox{  .}
\end{equation}
We impose $A_1=0$, since we assume that there is no evanescent inward component coming from the surface of the star. In the transition layer $(r_d\le r\le r_b-L_p)$, the general solution reads
\begin{equation}
\vec{z}=C_1\vec{z}_1^t+C_2\vec{z}_2^t \mbox{  ,}
\end{equation}
and in the radiative interior $(r\le r_d)$,
\begin{equation}
\vec{z}=B_1\vec{z}_1^r+B_2\vec{z}_2^r\mbox{  .}
\end{equation}
We choose $B_2=0$ because we only consider the inward component of the wave (travelling towards the center). In other words, we suppose that the waves will be damped before being reflected in the core, so that the emergence of modes is not allowed.

\setcounter{subsubsection}{0}
\subsection{Continuity of the wave function}
\label{continuity of the wave function}

To derive the wave energy flux at the top of the radiative zone $(r=r_d)$, we need to determine the modulus of the constant $B_1$. By imposing the continuity of the radial and the horizontal displacement (which is equivalent to continuity of the perturbation of pressure) at the points $r_p=r_b-L_p$ and $r_d$, we get four linear equations linking $B_1$ to $A_2$, $C_1$ and $C_2$, so that we close the system.

At the point $r=r_p$, corresponding to $z^\prime=d/2$ in \eqref{def N^2}, the argument of the Airy functions is equal to 
\begin{equation} \chi_p=\chi(r_p)=\left(\frac{k_hd\omega^2}{N_0^2 }\right) ^{2/3} << 1 \label{arg+d} \mbox{  ,}\end{equation} 
since $k_h d\le1$ and $\omega/N_0 \ll1$. Therefore, we can use the first-order Taylor expansion at $0$ of the functions $A_i(\chi)$ et $B_i(\chi)$ and their derivatives\footnote{Note that $A_i^\prime{}^\prime(0)=B_i^\prime{}^\prime(0)=0$, so that the first-order Taylor expansion of the derivatives of the Airy functions around $0$ is equal to their zero-order Taylor expansion.}
\begin{align}
v_{l,m}&=C_1^\prime+C_2^\prime \chi+\mathcal{O}\left(\chi^3\right) \label{taylor expansion}\\
\frac{d v_{l,m}}{d \chi}&=C_2^\prime+\mathcal{O}\left(\chi^2\right) \mbox{  ,}
\end{align}
where $C_1^\prime$ and $C_2^\prime$ are linked to $C_1$ and $C_2$ following the Taylor expansion of the Airy functions
\begin{align}
C_1&=\frac{3^{2/3}}{2} \Gamma(2/3)\Gamma(1/3)\left( \frac{C_1^\prime}{\Gamma(1/3)}-\frac{C_2^\prime}{3^{1/3}\Gamma(2/3)}\right) \label{C_1}\\
C_2&=\frac{3^{1/6}}{2} \Gamma(2/3)\Gamma(1/3)\left( \frac{C_1^\prime}{\Gamma(1/3)}+\frac{C_2^\prime}{3^{1/3}\Gamma(2/3)}\right)
\mbox{  .} \label{C_2}
\end{align} 
The continuity at the radius $r_p$ between the adiabatic region to the transition layer yields to
\begin{equation}
\mu_p \vec{z}_1^a(r_p)+\lambda_p \vec{z}_2^a(r_p)+A_2\vec{z}_2^a(r_p)=C_1 \vec{z}_1^t(r_p)+C_2\vec{z}_2^t(r_p) \mbox{  ,}
\label{continuity system at r_b}
\end{equation}
where $\mu_p=\mu(r_p)$ and $\lambda_p=\lambda(r_p)$. The second line of the system \eqref{continuity system at r_b} gives 
\begin{equation}
C_2^\prime=\frac{k_h^{1/2}}{K_1^{1/3}}\left[ \mu_p-(\lambda_p+A_2)\right] \mbox{  .}
\label{C_2 prime}
\end{equation}
 Using \eqref{C_2 prime} in the first line of \eqref{continuity system at r_b}, we obtain
\begin{equation}
C_1^\prime=k_h^{-1/2}\left[ \mu_p\left( 1-\frac{k_h d \omega^2}{N_0^2}\right)+(\lambda_p+A_2)\left( 1+\frac{k_h d \omega^2}{N_0^2}\right)\right] 
\label{C_1 prime}
\end{equation}

At the point $r=r_d$ $(z^\prime=-d/2)$, the argument of the Airy functions is equal to 
\begin{equation}
\chi_d=\chi(r_d)=-\left(\frac{k_h d N_0}{\omega}\right)^{2/3}\frac{N_0^2-\omega^2}{N_0^2} \mbox{  ,}
\end{equation}
and the continuity condition between the top of the radiative zone and the transition layer at $r_d$ yields
\begin{equation}
B_1 \vec{z}_1^r(r_d)=C_1\vec{z}_1^t(r_d)+C_2\vec{z}_2^t(r_d) \mbox{  .}
\label{continuity r_d}
\end{equation}

\addtocounter{subsubsection}{1} 
\subsubsection*{\Alph{section}.\arabic{subsection}.\arabic{subsubsection} Case of a sharp transition}

In the case of a sharp transition, i.e $|\chi_d|\ll1$, we can use again the first order Taylor expansion of the Airy functions. The second line of the system \eqref{continuity r_d} gives directly $C_2^\prime$ as a function of $B_1$
\begin{equation}
C_2^\prime=i \frac{k_h^{1/2}}{K_1^{1/3}}\left( \frac{N_0^2}{\omega^2}-1\right)^{1/4} B_1 \mbox{  .}
\label{C_2 prime sharp}
\end{equation}
Then, using \eqref{C_2 prime sharp} and \eqref{C_2 prime}, we can deduce $A_2$
\begin{equation}
A_2=-iB_1\left(\frac{N_0^2}{\omega^2}-1 \right)^{1/4}+\mu_p-\lambda_p  \mbox{  .}
\label{A_2 sharp}
\end{equation}
Now, \eqref{C_1 prime} enables us to determine $C_1^\prime$ as a function of $B_1$ by replacing $A_2$ by \eqref{A_2 sharp} 
\begin{equation}
C_1^\prime=k_h^{-1/2}\left[ 2\mu_p-iB_1\left( \frac{N_0^2}{\omega^2}-1\right)^{1/4}\left( 1+\frac{k_h d \omega^2}{N_0^2}\right)\right] \mbox{  .}
\label{C_1 prime sharp}
\end{equation}
 To go further, $C_1^\prime$ and $C_2^\prime$ are replaced by \eqref{C_1 prime sharp} and \eqref{C_2 prime sharp} in the first line of the system \eqref{continuity r_d} to obtain $B_1$
\begin{equation}
B_1=2\mu_p\left(\frac{N_0^2}{\omega^2}-1 \right)^{1/4}\left( 1+i\left(\frac{N_0^2}{\omega^2}-1 \right)^{1/2}(1+k_hd)\right)^{-1}
\mbox{  .}
\end{equation}
In the case of a sharp transition, we have $k_h d \ll 1$, and the modulus square of $B_1$ is finally given by
\begin {equation}
|B_1|^2\approx 4|\mu_p|^2\left(\frac{N_0^2}{\omega^2}-1\right)^{1/2}\frac{\omega^2}{N_0^2} \mbox{  .}
 \label{modulus B_1 sharp}
\end{equation}

\addtocounter{subsubsection}{1} 
\subsubsection*{\Alph{section}.\arabic{subsection}.\arabic{subsubsection} Case of a smooth transition}

We focus now on the case of a smooth transition, i.e., $|\chi_d|\gg 1$,
which is equivalent to $k_r d\gg 1$. We use the asymptotic expansion of the Airy functions and the one of their derivatives for $x \rightarrow+\infty$
\begin{align}
A_i(-x)&\sim x^{-1/2}  B_i^\prime(-x) \sim \frac{x^{-1/4}}{\sqrt{\pi}}\sin \left( \frac{2}{3}x^{3/2} +\frac{\pi}{4}\right) \label{A_i asymptote} \\
B_i(-x)&\sim -x^{-1/2}  A_i^\prime(-x) \sim \frac{x^{-1/4}}{\sqrt{\pi}}\cos\left( \frac{2}{3}x^{3/2} +\frac{\pi}{4}\right) \mbox{  .} \label{B_i asymptote}
\end{align}
Using \eqref{A_i asymptote}, \eqref{B_i asymptote} and the relation $|k_r|=K_1^{1/3}|\chi|^{1/2}$, \eqref{continuity r_d} can be rewritten as
\begin{align}
B_1&=K_1^{1/6}|\chi_d|^{1/4}\left[C_1 A_i(\chi_d)+C_2 B_i(\chi_d)\right]\label{B_r)}\\
iB_1&=K_1^{1/6}|\chi_d|^{1/4}\left[-C_1 B_i(\chi_d)+C_2 A_i(\chi_d)\right] \mbox{  .} \label{B_h}
\end{align}
After some algebra, we  find
\begin{align}
B_1&=K_1^{1/6}|\chi_d|^{-1/4} \frac{C_2}{\pi}\left[ B_i(\chi_d)+iA_i(\chi_d)\right]^{-1}\label{B f(C_2)}\\
B_1&=K_1^{1/6}|\chi_d|^{-1/4} \frac{C_1}{\pi}\left[ A_i(\chi_d)-iB_i(\chi_d)\right]^{-1} \mbox{  .}\label{B f(C_1)}
\end{align}
By computing the modulus of \eqref{B f(C_2)} and \eqref{B f(C_1)}, and dividing \eqref{B f(C_2)} by \eqref{B f(C_1)} , we obtain
\begin{align}
|B_1|^2&=K_1^{1/3}|C_1|^2=K_1^{1/3}|C_2|^2\label{modB vs modC_1}\\
\frac{C_2}{C_1}&=+i \mbox{  .}
\end{align}
To go further, we inject \eqref{C_1 prime} and \eqref{C_2 prime} into \eqref{C_1} and \eqref{C_2}, and we calculate \footnote{We used the equality $\Gamma(1/3) \Gamma(2/3)=2\pi/\sqrt{3}$}
\begin{align}
-C_1+\sqrt{3}C_2&=-2C_1e^{-i\pi/3}\nonumber\\
&=\frac{2 \pi}{3^{1/6}\Gamma(2/3)}\frac{k_h^{1/2}}{K_1^{1/3}}\left\{\mu_p-(\lambda_p+A_2)\right\}\label{C_1 vs A}\\
C_1+\sqrt{3}C_2&=2C_1e^{+i\pi/3}\nonumber\\
=\frac{2 \pi 3^{1/6}}{\Gamma(1/3)}k_h^{-1/2}&\left\{\mu_p\left( 1-\frac{k_h d \omega^2}{N_0^2}\right)+(\lambda_p+A_2)\left( 1+\frac{k_h d \omega^2}{N_0^2}\right)\right\} \mbox{  .}\label{C_2 vs A}
\end{align}
Using \eqref{C_1 vs A}, we get $A_2$ as a function of $C_1$ that we reinject into \eqref{C_2 vs A} to find
\begin{equation}
C_1e^{+i\pi/3}\left[1-3^{1/3}e^{-2i\pi/3}\frac{\Gamma(2/3)}{\Gamma(1/3)}\frac{K_1^{1/3}}{k_h}(1+ \frac{k_h d \omega^2}{N_0^2})\right]
=\frac{2 \pi 3^{1/6}}{\Gamma(1/3)}k_h^{-1/2} \mu_p \label{C_1 final}
\end{equation}
Finally, we can assume
\begin{equation}
\frac{K_1^{1/3}}{k_h}=\left(\frac{ N_0^2 }{k_h d \omega^2}\right)^{1/3}>>1 \mbox{  ,}
\end{equation}
and we compute the absolute square of \eqref{C_1 final} to obtain, using \eqref{modB vs modC_1},
\begin{equation}
|B_1|^2\approx\left( \frac{3^{-1/6}\pi}{\Gamma(2/3)}\right)^2 4|\mu_p|^2 \left( \frac{k_h d \omega^2}{N_0^2}\right)^{1/3} \mbox{  .}
\label{modulus B_1 smooth}
\end{equation}
Thus, we find a similar expression for $|B_1|^2$ to what we obtained in the case of a sharp transition. They differ from each other according to a structural factor, corresponding to the ratio of the transmission coefficients in both cases,
\begin{align}
\left(|B_1|^2\right)_{smooth}&= D  \left( \frac{k_h d N_0}{\omega}\right)^{1/3} \left(\frac{N_0^2}{N_0^2-\omega^2}\right)^{1/2}\left(|B_1|^2\right)_{sharp} \nonumber \\
&\sim D \left( k_r d \right)^{1/3} \left(|B_1|^2\right)_{sharp} \mbox{  ,}
\label{B_1 smooth vs sharp}
\end{align}
where $D$ is a numerical factor
\begin{equation}
D=\left( \frac{3^{-1/6}\pi}{\Gamma(2/3)}\right)^2\sim 3.7 \mbox{  .}
\end{equation}
%

\subsection{Final expression for the wave energy flux}
\label{final expression of the WEF}

In the radiative zone, the wave function is equal to $B_1 \vec{z}_1^r$. Then, using \eqref{y_1^r}, \eqref{spectral density term} becomes
\begin{align}
\widetilde{\mathcal{E}}_{l,m}(r,\omega)&= \frac{\nu_p}{8 \pi^2} \rho \int_{\Omega_0} \left\{ \left|\hat{\mathrm{v}}_{r,l,m}\right|^2+l(l+1) \left|\hat{\mathrm{v}}_{h,l,m}\right|^2 
\right\} \frac{\mathrm{d}\Omega_0}{4\pi} \mbox{  ,}\\
&=\frac{\nu_p}{8 \pi^2}\frac{\sqrt{l(l+1}}{r^3}\frac{N^2}{\omega^2}\left(\frac{N^2}{\omega^2}-1 \right)^{-1/2}\int_{\Omega_0}|B_1|^2
\frac{\mathrm{d}\Omega_0}{4\pi} \mbox{  .}
\end{align}
Given \eqref{flux (l,m)} and \eqref{group_velocity}, the mean radial wave energy flux reads
\begin{equation}
\mathcal{F}_r(r,\omega,l,m)=\mathcal{N} \frac{\nu_p}{8 \pi^2}\frac{\omega}{r^2}\int_{\Omega_0} |B_1|^2\frac{\mathrm{d}\Omega_0}{4\pi} \mbox{  .}
\label{wave flux general}
\end{equation}
In the case of a sharp transition, using \eqref{modulus B_1 sharp}, \eqref{coeff mu}, \eqref{alpha omega} and \eqref{beta l_m annexe}, \eqref{wave flux general} becomes
\begin{equation}
\mathcal{F}_r(r,\omega,l,m)=\frac{\mathcal{N}}{4} \frac{\sqrt{l(l+1}}{4 \pi r^2}\frac{\left(N_0^2-\omega^2\right)^{1/2}}{N_0^2} \frac{e^{-\omega^2/4 \nu_p^2}}{\nu_p} \mathcal{B}_l^m \mathcal{H}_l^2 \mbox{  ,}
\end{equation}
where we have defined
\begin{align}
\mathcal{B}_l^m&=\frac{1}{4 \pi} \int \left| \beta_l^m\right|^2\sin \theta_0 \mathrm{d} \theta_0 \mathrm{d} \phi_0 \\
\mathcal{H}_l&= \int_{r_b-L_p}^{r_b} \frac{d}{dr}\left( \rho V_0^2\right)
\rho^{-1/2} r^{1/2} \left( \frac{r_b-L_p}{r}\right)^{\Lambda}  \mathrm{d} r \mbox{  .}
\end{align}
In the case of a smoother transition, the derivation is similar and simple if we use \eqref{B_1 smooth vs sharp}, so that we obtain
\begin{equation}
\mathcal{F}_{r,smooth}=D  \left( \frac{k_h d N_0}{\omega}\right)^{1/3} \left(\frac{N_0^2}{N_0^2-\omega^2}\right)^{1/2}\mathcal{F}_{r,sharp}
\mbox{  .}
\end{equation}

\end{document}